\pdfoutput=1
   \documentclass[fleqn,usenatbib,usedcolumn]{mnras}
   \usepackage[british]{babel}             
   \usepackage{newtxtext}                  
   \usepackage[slantedGreek]{newtxmath}    
   \let\la=\lesssim 
   \usepackage[T1]{fontenc}                

\usepackage{graphicx,color}

\def\hi{\mbox{H\sc{i}}}
\def\nii{\mbox{N\sc{ii}}}

\def\kms{km\,s$^{-1}$}

\def\msun{M$_{\odot}$}

\def\arcsec{$^{\prime \prime}$}
\definecolor{Mygrey}{gray}{0.75}

\newcommand{\ltsimeq}{\raisebox{-0.6ex}{$\,\stackrel{\raisebox{-.2ex}{$\textstyle <$}}{\sim}\,$}}
\newcommand{\gtsimeq}{\raisebox{-0.6ex}{$\,\stackrel{\raisebox{-.2ex}{$\textstyle >$}}{\sim}\,$}}
\newcommand{\farc}{\mbox{\ensuremath{.\!\!^{\prime\prime}}}}
\newcommand{\fdeg}{\mbox{\ensuremath{.\!\!^{\circ}}}}
\newcommand{\coordsec}{\mbox{\ensuremath{.\!\!^{\rm s}}}}
\mathchardef\mhyphen="2D

\usepackage[compact]{titlesec}
\titlespacing{\section}{0pt}{*2}{*1}

   \hypersetup{pdfauthor={T. A. Davis},
               pdftitle={Revealing the intermediate-mass black hole at the heart of the dwarf galaxy NGC\,404 with sub-parsec resolution ALMA observations},
               pdfkeywords={galaxies: individual: NGC\,404 -- galaxies: elliptical and lenticular, cD -- galaxies: dwarf -- galaxies: ISM -- galaxies: evolution -- galaxies: kinematics and dynamics},
               bookmarksnumbered=true}

\title[ALMA reveals an IMBH in NGC\,404]{Revealing the intermediate-mass black hole at the heart of the dwarf galaxy NGC\,404 with sub-parsec resolution ALMA observations} 
\author[Timothy A. Davis et al.]{\parbox{\textwidth}{Timothy A. Davis,$^{1}$\thanks{E-mail: \texttt{DavisT@cardiff.ac.uk}} Dieu D. Nguyen,$^{2}$ Anil C. Seth,$^{3}$ Jenny E. Greene,$^{4}$ Kristina Nyland,$^{5}$ Aaron J. Barth,$^{6}$ Martin Bureau,$^{7,8}$ Michele Cappellari,$^{7}$ Mark den Brok,$^{9}$ Satoru Iguchi,$^{2,10}$ Federico Lelli,$^{1}$ Lijie Liu,$^{7}$ Nadine Neumayer,$^{11}$ Eve V. North,$^{1}$ Kyoko Onishi,$^{12}$ Marc Sarzi,$^{13}$ Mark D. Smith$^{7}$ and Thomas G. Williams$^{11}$ }
\vspace{0.4cm}\\
\parbox{\textwidth}{$^{1}$School of Physics \&\ Astronomy, Cardiff University, Queens Buildings, The Parade, Cardiff, CF24 3AA, UK\\
$^{2}$National Astronomical Observatory of Japan (NAOJ), National Institute of Natural Sciences (NINS), 2-21-1 Osawa, Mitaka, Tokyo
181-8588, Japan\\
$^{3}$Department of Physics and Astronomy, University of Utah, 115 South 1400 East, Salt Lake City, UT 84112, USA\\
$^{4}$Department of Astrophysics, Princeton University, Princeton, NJ 08544, USA\\
$^{5}$National Research Council, Resident at the U.S. Naval Research Laboratory, 4555 Overlook Ave SW, Washington, DC, 20375, USA\\
$^{6}$Department of Physics and Astronomy, University of California, Irvine, 4129 Frederick Reines Hall, Irvine, CA 92697, USA\\
$^{7}$Sub-department of Astrophysics, Department of Physics, University of Oxford, Denys Wilkinson Building, Keble Road, Oxford OX1 3RH, UK\\
$^{8}$Yonsei Frontier Lab and Department of Astronomy, Yonsei University, 50 Yonsei-ro, Seodaemun-gu, Seoul 03722, Republic of Korea\\
$^{9}$Leibniz-Institut f\"ur Astrophysik Potsdam (AIP), An der Sternwarte 16, 14482 Potsdam, Germany\\
$^{10}$Department of Astronomical Science, SOKENDAI (The Graduate University of Advanced Studies), Mitaka, Tokyo 181-8588, Japan\\
$^{11}$Max-Planck-Institut f\"ur Astronomie, K\"onigstuhl 17, 69117 Heidelberg, Germany\\
$^{12}$Department of Space, Earth and Environment, Chalmers University of Technology, Onsala Observatory, SE-439 92 Onsala, Sweden\\
$^{13}$Armagh Observatory and Planetarium, College Hill, Armagh BT61 9DG, UK
}}
 \volume{{\rm in press}}
\begin{document}

\date{Accepted 2020 May 29. Received 2020 May 25; in original form 2020 April 1}

\pagerange{\pageref{firstpage}--\pageref{lastpage}} \pubyear{2020}

\maketitle

\label{firstpage}

\begin{abstract}
We estimate the mass of the intermediate-mass black hole at the heart of the dwarf elliptical galaxy NGC\,404 using Atacama Large Millimeter/submillimeter Array (ALMA) observations of the molecular interstellar medium at an unprecedented linear resolution of $\approx$0.5\,pc, in combination with existing stellar kinematic information.  These ALMA observations reveal a central disc/torus of molecular gas clearly rotating around the black hole. This disc is surrounded by a morphologically and kinematically complex flocculent distribution of molecular clouds, that we resolve in detail. Continuum emission is detected from the central parts of NGC\,404, likely arising from the Rayleigh--Jeans tail of emission from dust around the nucleus, and potentially from dusty massive star-forming clumps at discrete locations in the disc. Several dynamical measurements of the black hole mass in this system have been made in the past, but they do not agree. We show here that both the observed molecular gas and stellar kinematics independently require a $\approx5\times10^5$\,\msun\ black hole once we include the contribution of the molecular gas to the potential.
Our best estimate comes from the high-resolution molecular gas kinematics, suggesting the black hole mass of this system is 5.5$^{+4.1}_{-3.8}\times$10$^5$ \msun\ (at the 99\% confidence level), in good agreement with our revised stellar kinematic measurement and broadly consistent with extrapolations from the black hole mass -- velocity dispersion and black hole mass -- bulge mass relations. 
 This highlights the need to accurately determine the mass and distribution of each dynamically important component around intermediate-mass black holes when attempting to estimate their masses.
\end{abstract}

\begin{keywords}
galaxies: individual: NGC\,404 -- galaxies: elliptical and lenticular, cD -- galaxies: dwarf -- galaxies: ISM -- galaxies: evolution -- galaxies: kinematics and dynamics
\end{keywords}

\renewcommand{\thefootnote}{\textsuperscript{\arabic{footnote}}}
\section{Introduction}

Supermassive black holes (SMBHs) are ubiquitous at the centres of massive galaxies. Over the past few decades the masses of these SMBHs have been carefully measured using stellar and gas dynamical methods (see e.g. the review by \citealt{2013ARA&A..51..511K}). These measurements reveal that SMBH masses scale with many properties of their host galaxies. However, much less is known about the presence of black holes (BHs) in dwarf galaxies, let alone whether they obey the scaling relations observed at higher masses \citep[e.g.][]{2012NatCo...3E1304G}.

Given the difficulty of growing SMBHs from stellar mass BHs (even when accreting for a Hubble time), it is thought that massive BHs must instead grow from more massive seeds (e.g. \citealt{2001ApJ...551L..27M,2003ApJ...582..559V,2019arXiv191105791I,2019arXiv191109678G}). The exact processes that form these seeds, and the seeds' intrinsic mass distribution, are currently unknown.  Measurements of the masses of central BHs in dwarf galaxies can constrain these important, but very poorly understood processes. While in massive galaxies the memory of the original BH has long since been erased, dwarf galaxies experience far less merging or accretion. The mass distribution, occupation fraction and scaling relations of the intermediate-mass black holes (IMBHs) in dwarf galaxies today can thus help constrain seeding mechanisms (e.g. \citealt{2012Sci...337..544V, 2018MNRAS.481.3278R}).

{Some dwarf galaxies with stellar masses $\sim$10$^9$~M$_\odot$ appear not to host a central BH. In the Local Group, \citet{2001AJ....122.2469G} published an upper limit of just 1500 M$_\odot$ for the mass of a putative BH in the low-mass spiral galaxy M33, while in the dwarf elliptical galaxy NGC~205 any central BH has a mass below $\sim$10$^4$~M$_\odot$ \citep{2005ApJ...628..137V, 2019ApJ...872..104N}.} The four closest dwarf galaxies outside the Local Group known to host a central BH are NGC\,404, NGC\,4395, NGC\,5102, and NGC\,5206 and all have published dynamical BH mass estimates \citep{2010ApJ...714..713S, 2015ApJ...809..101D,2017ApJ...836..237N,2018ApJ...858..118N,2019ApJ...872..104N}. However, weighing these BHs is at the edge of what is feasible with state-of-the-art adaptive optics observations. Thus, dynamical measurements of BH masses in dwarf galaxies (expected to have BH masses $<$10$^6$ M$_{\odot}$) do not extend much beyond the Local Group, limiting detection of these BHs to the small subset of objects that are actively accreting \citep[e.g.][]{2008ApJ...688..159G,2013ApJ...775..116R,2015ApJ...799...98M}.

In this paper we report on sub-parsec resolution Atacama Large Millimeter/submillimeter Array (ALMA) molecular gas observations of the dwarf elliptical galaxy NGC\,404. This object has a total stellar mass of $\approx1.2\times10^9$\,\msun\ \citep{2010ApJ...714..713S},  hosts a large-scale low-surface brightness region of star formation (coincident with a large scale \hi\ ring) and a disc/ring of dust closer to its centre, both likely accreted during a recent (minor) merger \citep{2004AJ....128...89D,2010ApJ...714L.171T}. NGC\,404 also hosts an accreting BH (see e.g. \citealt{2017ApJ...845...50N}).  Limits on the mass of this BH have been estimated \citep{2010ApJ...714..713S, 2017ApJ...836..237N} both from modelling of the stellar kinematics (yielding {M$_{\rm BH}<1.5\times10^5$\,\msun}), and using the kinematics of a rotating disc of hot ($\approx$2300\,K) molecular hydrogen at its centre (suggesting {M$_{\rm BH}<2\times10^5$\,\msun}). The best-fitting measurements from these papers do not agree particularly well, perhaps because the hot-H$_2$ gas has significant non-circular motions. However, we also know that this object has a subtantial mass of molecular gas at its centre ($\approx$9$\times10^6$\,\msun; \citealt{2015AJ....149..187T}), that may contribute significantly to the galaxy potential and thus bias the existing BH mass measurements. Depending on the range of radii where the gas is dynamically important, this could lead to either an over- or under-estimate of the BH mass.  Here we probe the distribution of the cold CO(2-1) emitting molecular gas in NGC\,404 at sub-parsec scales with ALMA. With these data (which reach higher spatial and spectral resolution than all previous observations) we not only obtain an independent estimate of the BH mass using the kinematics of the cold molecular material \cite[e.g.][]{2013Natur.494..328D}, but can also improve our stellar kinematic estimates of the BH mass in this object. 

In Section \ref{data} we discuss the ALMA data this paper is based on. In Section \ref{stellar_model} we incorporate the molecular gas distribution into our stellar kinematic models, and derive a new constraint on the BH mass of NGC\,404. In Section \ref{model} we model the kinematics of the molecular gas itself and obtain an independent estimate of the BH mass. We discuss our results in Section \ref{discuss} before concluding in Section \ref{conclude}.  We assume a distance of 3.06 $\pm$ 0.37 Mpc to NGC\,404 \citep[derived using the tip of the red giant branch method;][]{2002A&A...389..812K}, yielding a physical scale of 14.8 pc arcsec$^{-1}$.

 \begin{figure*}
\begin{center}
\includegraphics[width=0.7\textwidth,angle=0,clip,trim=0.0cm 0.0cm 0cm 0.0cm]{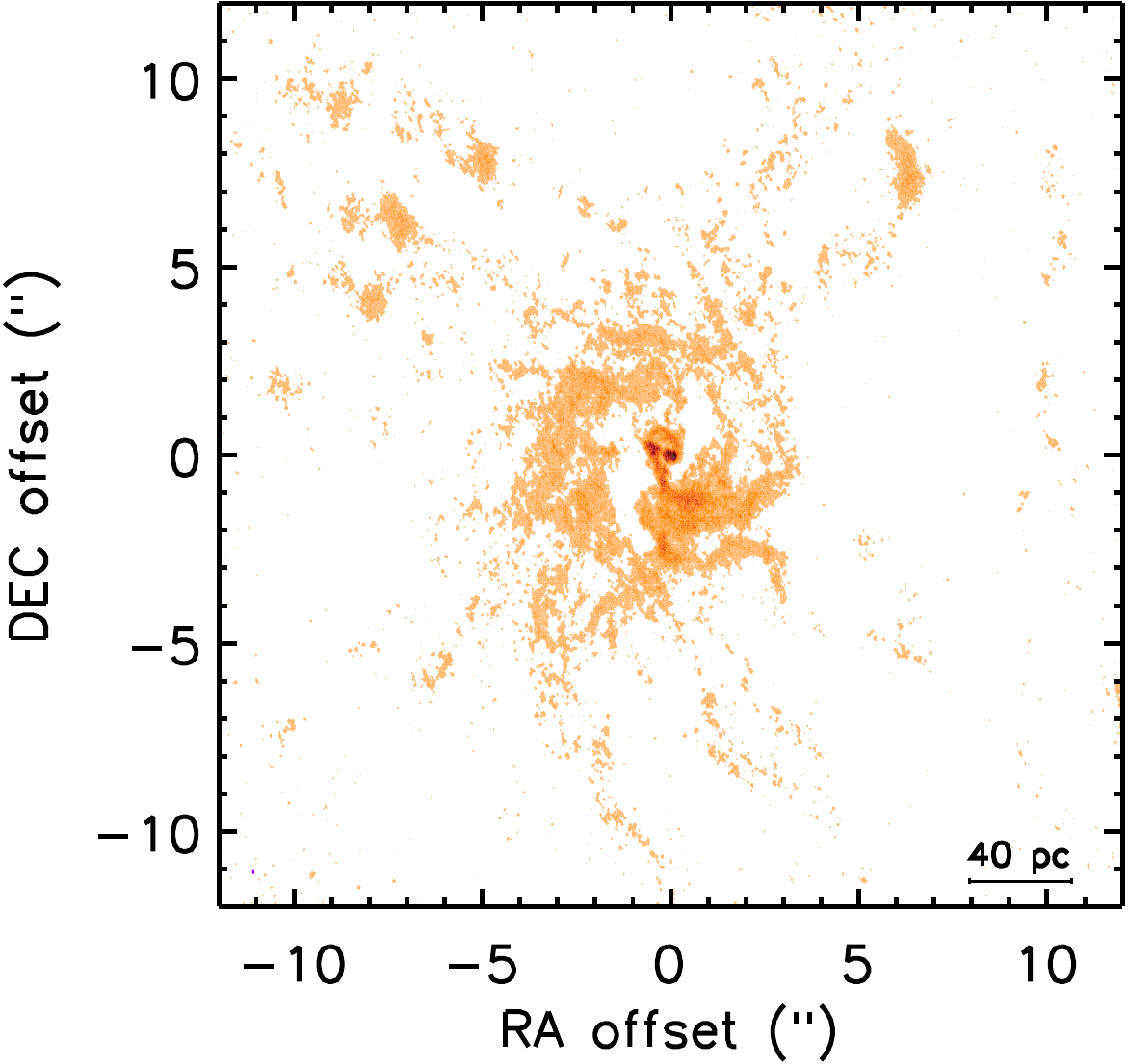}
\caption{Zeroth moment (integrated intensity) map of the $^{12}$CO(2-1) emission in NGC\,404, created from our combined dataset using the masked-moment technique described in Section \ref{data}. The synthesised beam (0\farc078\,$\times$\,0\farc037 or $\approx$0.8 pc$^2$) is shown as an extremely small {purple} ellipse at the bottom left corner of the figure. Our $\approx$0.8 pc resolution data reveal the complex morphology of the gas in the centre of this gas-rich dwarf elliptical galaxy.}
\label{ALMAmoments}
 \end{center}
 \end{figure*}
 
  \begin{figure*}
\begin{center}
\includegraphics[height=6.7cm,angle=0,clip,trim=0cm 0.0cm 0cm 0.0cm]{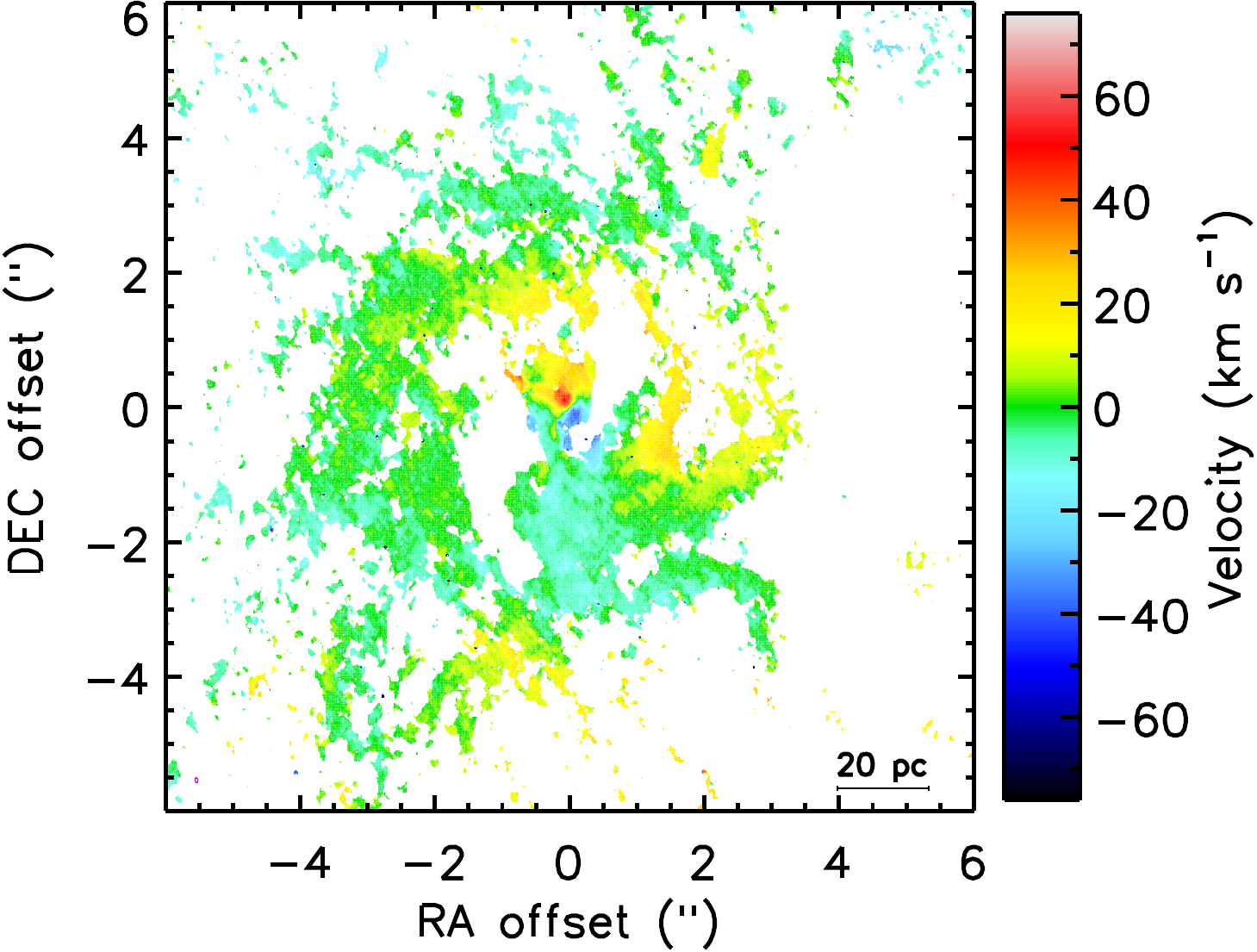}\hspace{0.25cm}
\includegraphics[height=6.7cm,angle=0,clip,trim=1.1cm 0.0cm 0cm 0.0cm]{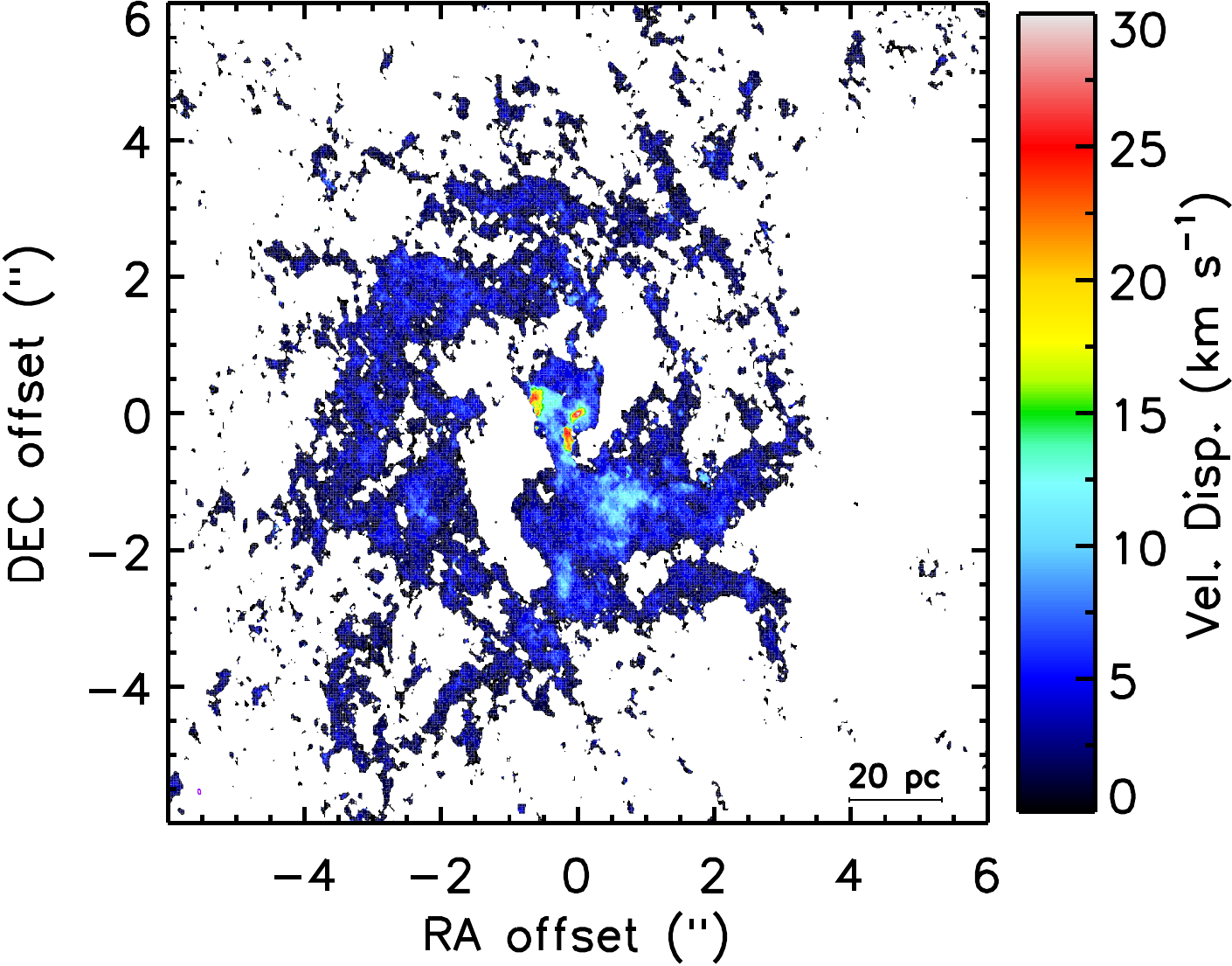}
\caption{First and second moment (mean line-of-sight velocity and velocity dispersion) maps of the $^{12}$CO(2-1) emission in NGC\,404, created from our combined dataset using the masked-moment technique described in Section \ref{data}. The synthesised beam (0\farc078\,$\times$\,0\farc037 or $\approx$0.8 pc$^2$) is shown as an extremely small {purple} ellipse at the bottom left of each panel. In the left panel, the velocity is shown relative to the systemic velocity of NGC\,404, here assumed to be $-53$ \kms. The gas in NGC\,404 is rotating, at least in the centre, but becomes kinematically complex in the outer parts. }
\label{ALMAkinmoments}
 \end{center}
 \end{figure*}

\section{ALMA data}
\label{data}
The  $^{12}$CO(2--1) line in NGC\,404 was observed with an extended ALMA configuration twice on 31st October 2015, as part of programme 2015.1.00597.S.  These data, which detected the central disc in this object, were presented in \cite{2017ApJ...845...50N}.  In this work we include three additional ALMA tracks. The first, taken with extended baselines on 5th September 2015 did not pass initial quality assurance due to the lack of a suitable amplitude calibrator, but upon further inspection was found to be usable. Two additional lower resolution tracks were obtained on 16th June and 3rd September 2018 as part of project 2017.1.00572.S.
  
In each of these observations an 1850 MHz correlator window was placed over the $^{12}$CO(2--1) line, yielding a continuous velocity coverage of $\approx$2000 \kms\ with a raw channel width of $\approx$0.6 \kms, sufficient to properly cover and sample the line. Three additional 2 GHz-wide low-resolution correlator windows were simultaneously used to probe continuum emission. 

The raw ALMA data were calibrated using manual calibration scripts in the \texttt{Common Astronomy Software Applications} {(\tt CASA)} package. 
Phase and bandpass calibration were performed using the quasars J2253+1608, J0237+2848, J0112+3522 and J0112+3208, while flux calibration was performed using quasar J0238+1636.

   \begin{figure*}
\begin{center}
\includegraphics[height=4.85cm,angle=0,clip,trim=0.0cm 0.0cm 0cm 0.0cm]{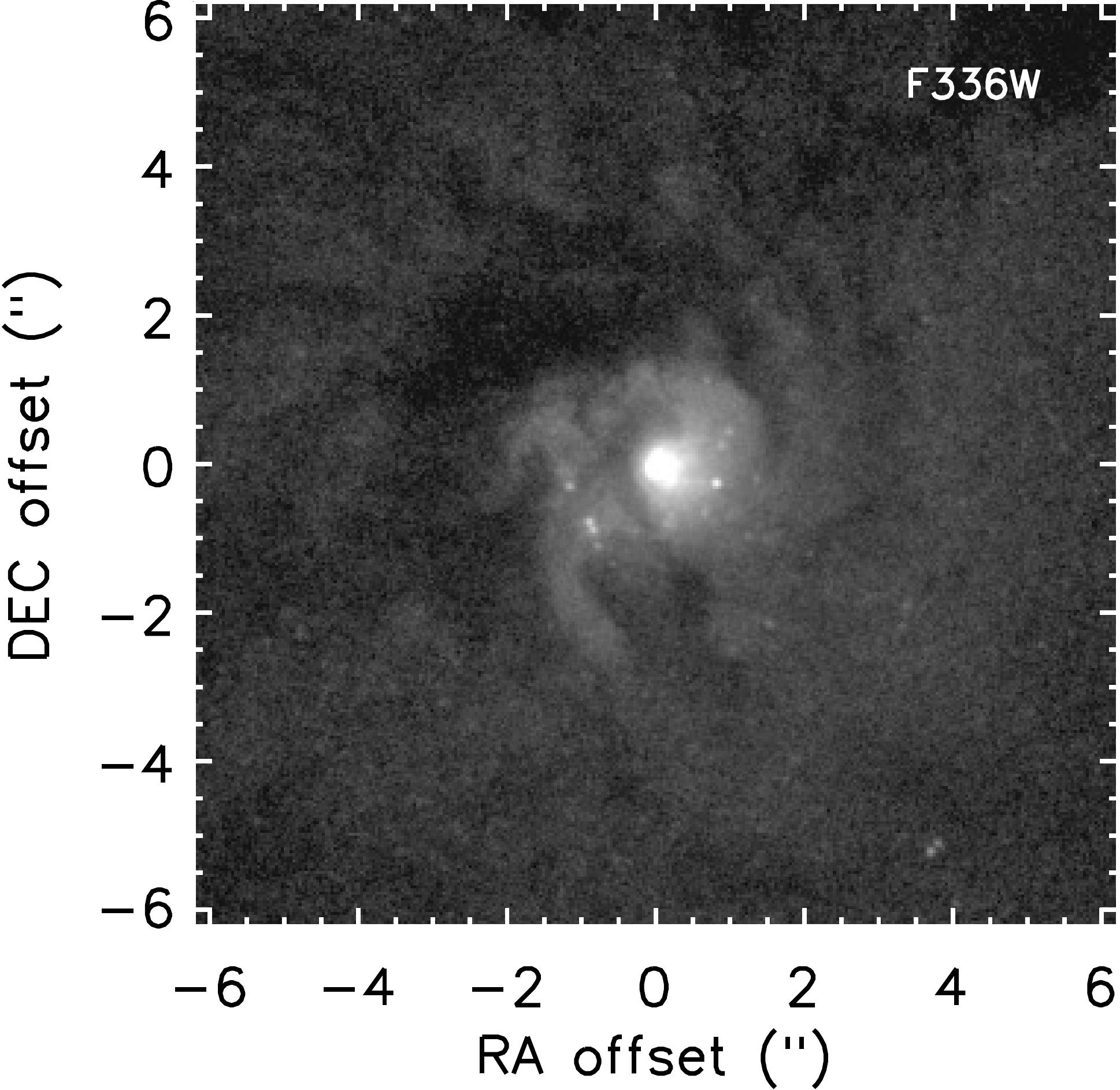}
\includegraphics[height=4.85cm,angle=0,clip,trim=3.2cm 0.0cm 0cm 0.0cm]{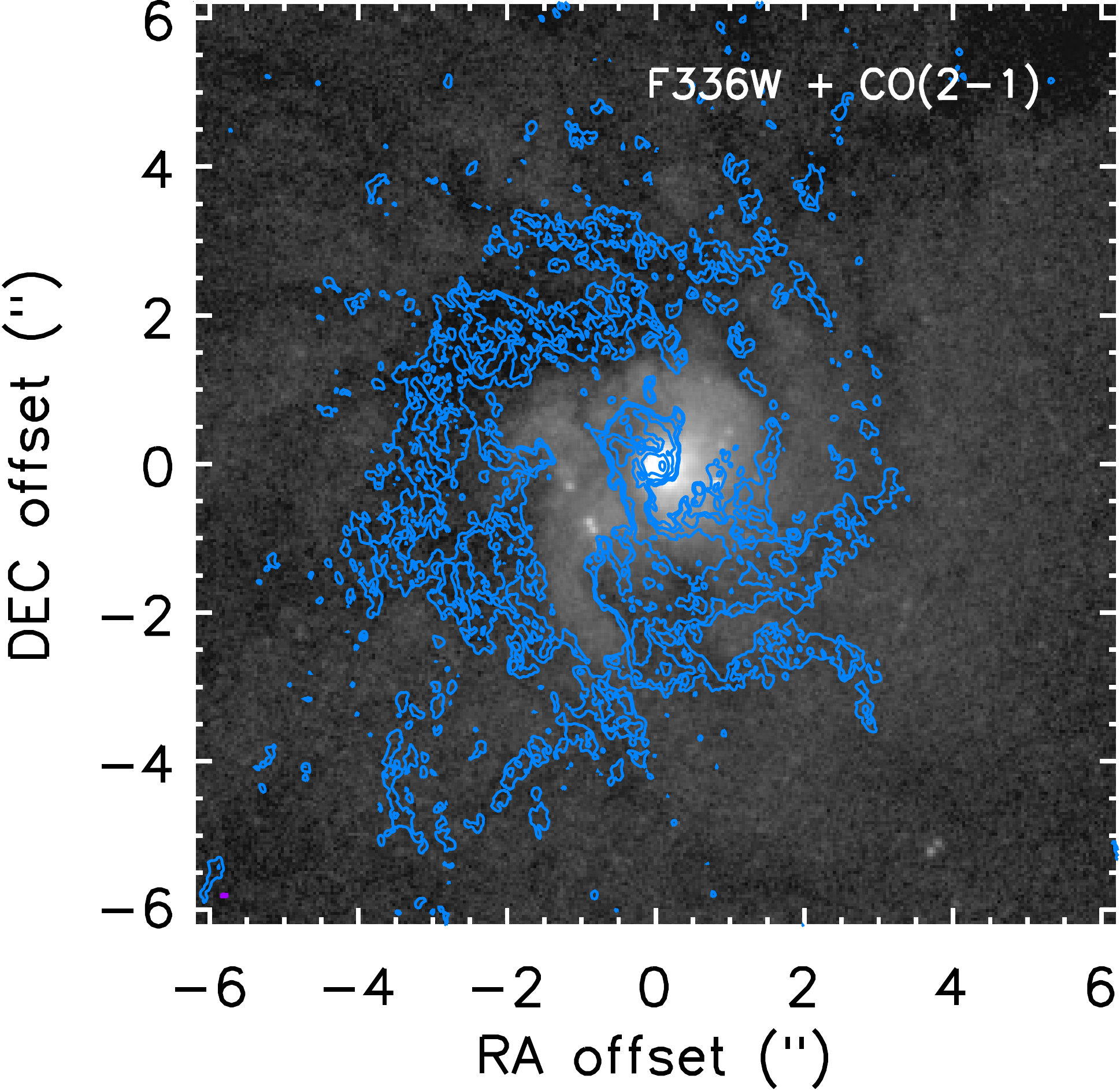}
\includegraphics[height=4.85cm,angle=0,clip,trim=3.2cm 0.0cm 0cm 0.0cm]{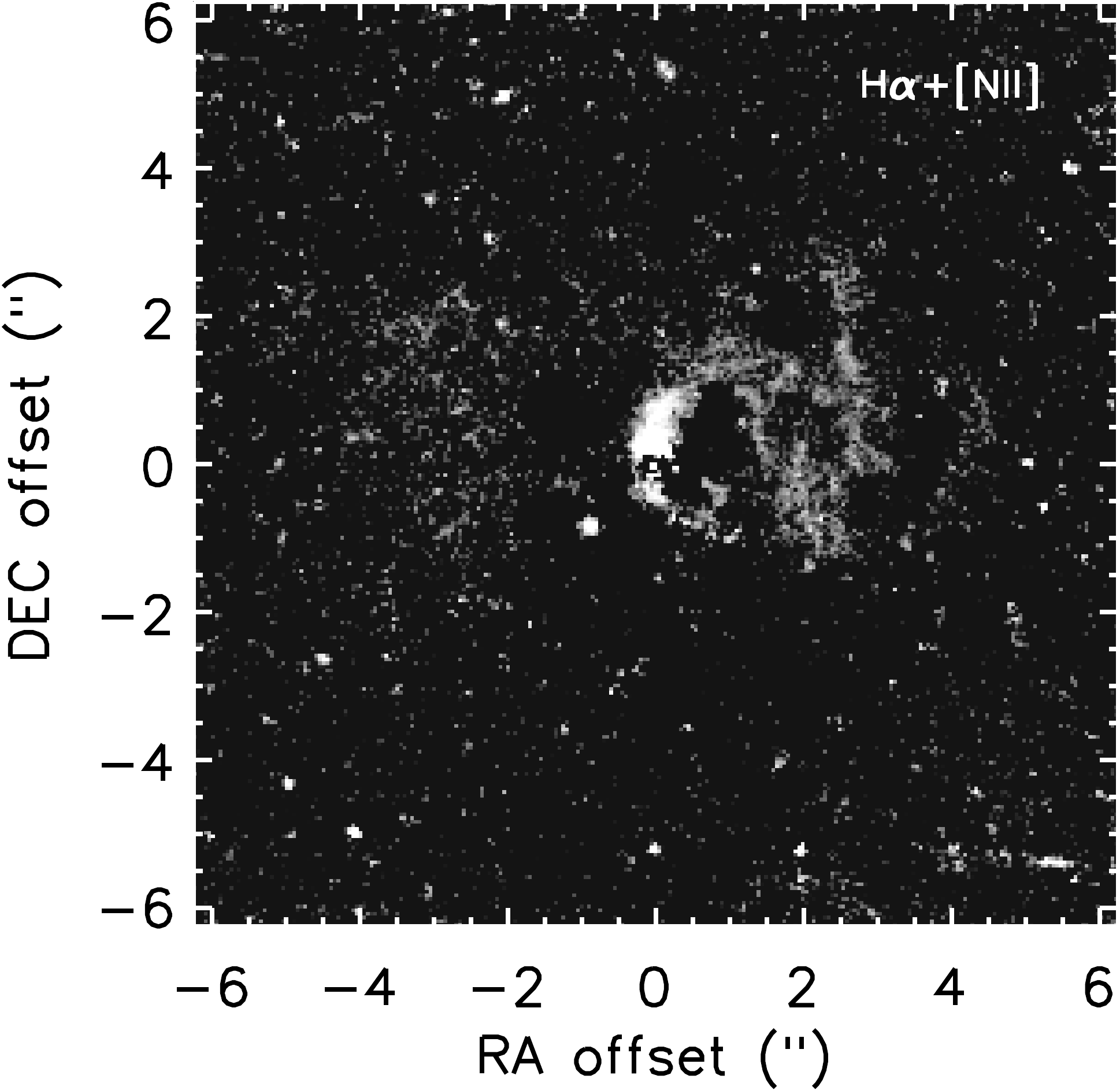}
\includegraphics[height=4.85cm,angle=0,clip,trim=3.2cm 0.0cm 0cm 0.0cm]{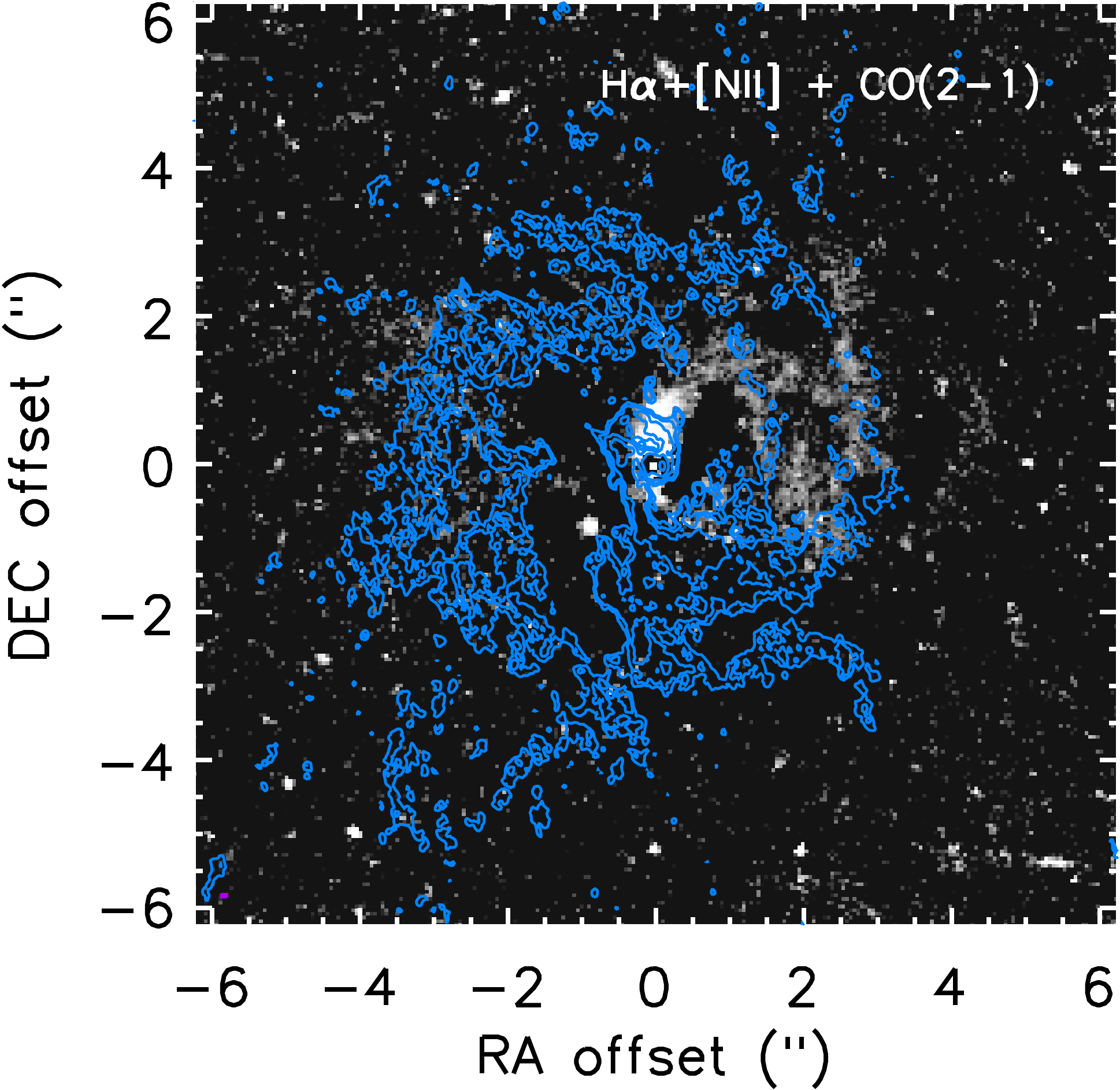}
\caption{\textit{Left panel:} Unsharp-masked \textit{HST} Wide Field Camera 3 (WFC3) F336W image of a 180 pc $\times$ 180 pc region around the nucleus of NGC\,404, revealing a clear central dust disc/ring. 
\textit{Centre-left panel:} As before, but overlaid with blue $^{12}$CO(2-1) integrated intensity contours from our combined ALMA dataset. The synthesised beam (0\farc078\,$\times$\,0\farc037 or $\approx$0.8 pc$^2$) is shown as a (very small) {purple} ellipse at the bottom left of the panel. The molecular gas disc coincides well with the dust structures seen in absorption.
\textit{Centre-right panel:} Narrow-band \textit{HST} image (F656N-F547M) of the H$\alpha$+[\nii] lines in NGC\,404. \textit{Right panel:} As before, but overplotted with blue $^{12}$CO(2-1) contours as in the central panel. Strong ionised-gas line emission is present in the centre and further out, where it fills some of the gaps in the molecular gas/dust ring.}
\label{ALMA_HST_over}
 \end{center}
 \end{figure*}
 
We used the {\tt CASA} package to combine and image the visibility files of the different tracks. For each of the data cubes discussed below, continuum emission was detected, and subtracted in the $uv$-plane using the \textsc{CASA} task \textit{uvcontsub}. These continuum subtracted data were then imaged in order to produce a three-dimensional RA-Dec-velocity data cube (with velocities determined with respect to the rest frequency of the $^{12}$CO(2-1) line).  
This dirty cube was cleaned in regions of source emission (identified interactively) to a threshold equal to twice the root-mean square (RMS) noise of the dirty channels. Our source is very extended at the spatial resolution of our observations, and thus we used a multi-scale clean algorithm \citep{2011A&A...532A..71R} with cleaning scales optimised for point sources, and emission with a characteristic scale 3, 6 and 9 times the synthesised beam. The clean components found from this analysis were then added back and re-convolved using a Gaussian beam of full-width-at-half-maximum (FWHM) equal to that of the dirty beam. The resulting cube was then corrected for the primary beam response. 

In this work we make use of two different sets of CO(2-1) data products. The first {high-resolution dataset}, which we use for our kinematic modelling, was produced from the three long baseline tracks only, imaged using Briggs weighting (robust parameter 0.5). This yields a synthesised beam of 0\farc051\,$\times$\,0\farc026 at a position angle of $-10^{\circ}$ (a mean physical resolution of 0.54 pc) and an RMS noise of 0.79 mJy beam$^{-1}$ channel$^{-1}$. This dataset allows us to image the central disc at the highest spatial resolution possible, but it does not recover all the flux. The second {combined dataset} was produced from all available ALMA tracks, imaged using Briggs weighting (robust parameter 0.1). This yields a synthesised beam of 0\farc078\,$\times$\,0\farc037 at a position angle of 21$^{\circ}$ (a mean physical resolution of 0.8 pc), and an RMS noise of 0.61 mJy beam$^{-1}$ channel$^{-1}$. This combination allows us to retrieve almost all the flux from this object (see below), but at a slight penalty to spatial resolution. We use this dataset when discussing the morphology of the gas.  Both datasets use a channel width of 2\,\kms\ and pixels of 0\farc015$\times$0\farc015.

 \begin{figure}
\begin{center}
\includegraphics[width=0.48\textwidth,angle=0,clip,trim=0.0cm 0.0cm 0cm 0.0cm]{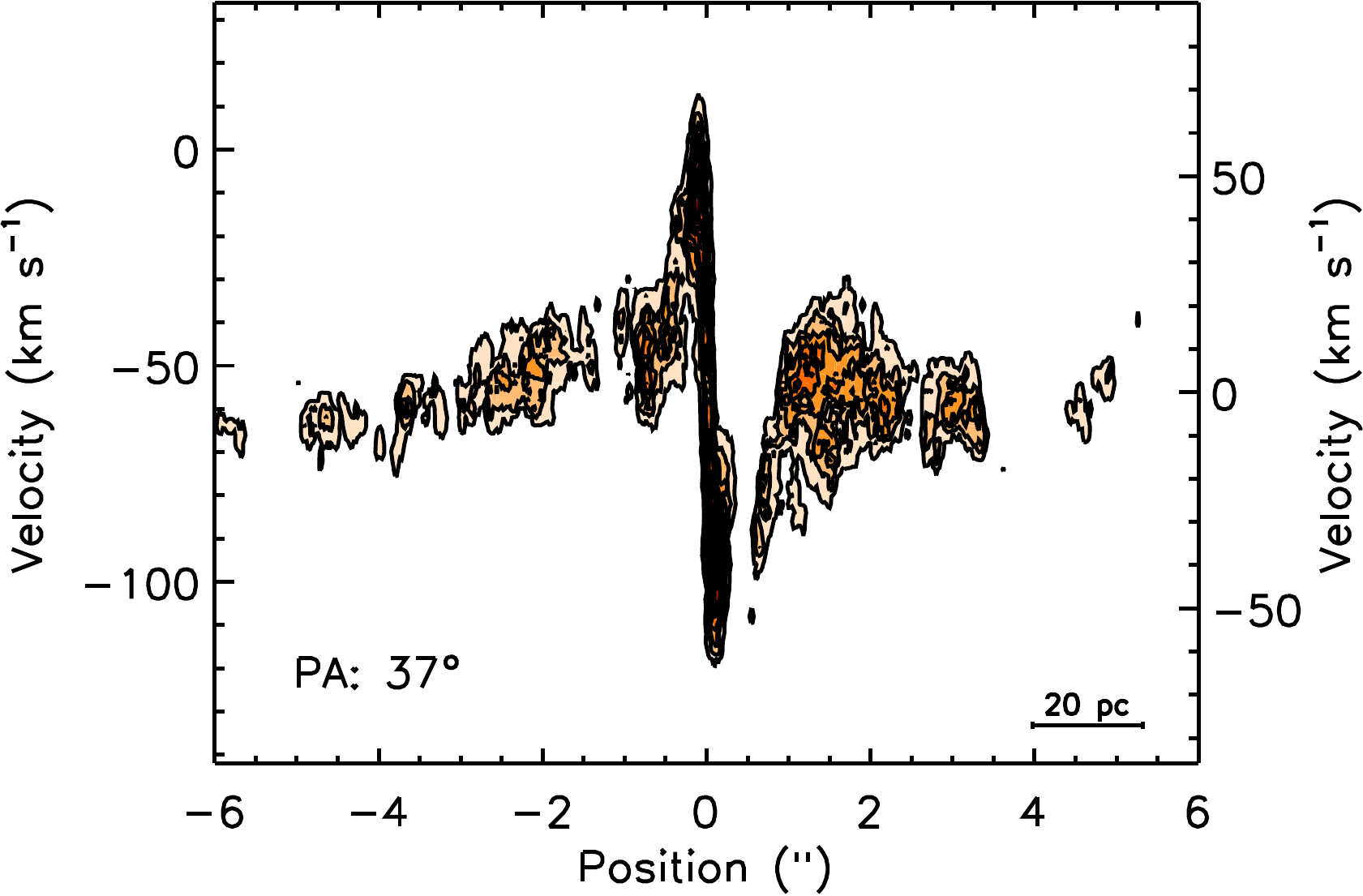}
\caption{Position-velocity diagram of the $^{12}$CO(2-1) emission in NGC\,404 from our combined dataset, extracted along the kinematic major axis of the central disc-like emission (37$^{\circ}$; see Section \ref{model}) with a slit of 5 pixels in width. We do not show the synthesised beam (0\farc078\,$\times$\,0\farc037 or $\approx$0.8 pc$^2$) or velocity channel width (2 \kms) explicitly in this plot, as they are very small compared to the ranges plotted. A clear signature of enhanced rotation is visible around the nuclear star cluster/putative IMBH, while the line-of-sight velocity of the gas is low in the outer parts, where the disc warps to become approximately face on.}
\label{ALMA_pvd}
 \end{center}
 \end{figure}

 \begin{figure*}
\begin{center}
\includegraphics[width=0.75\textwidth,angle=0,clip,trim=0cm 0.0cm 0cm 0.0cm]{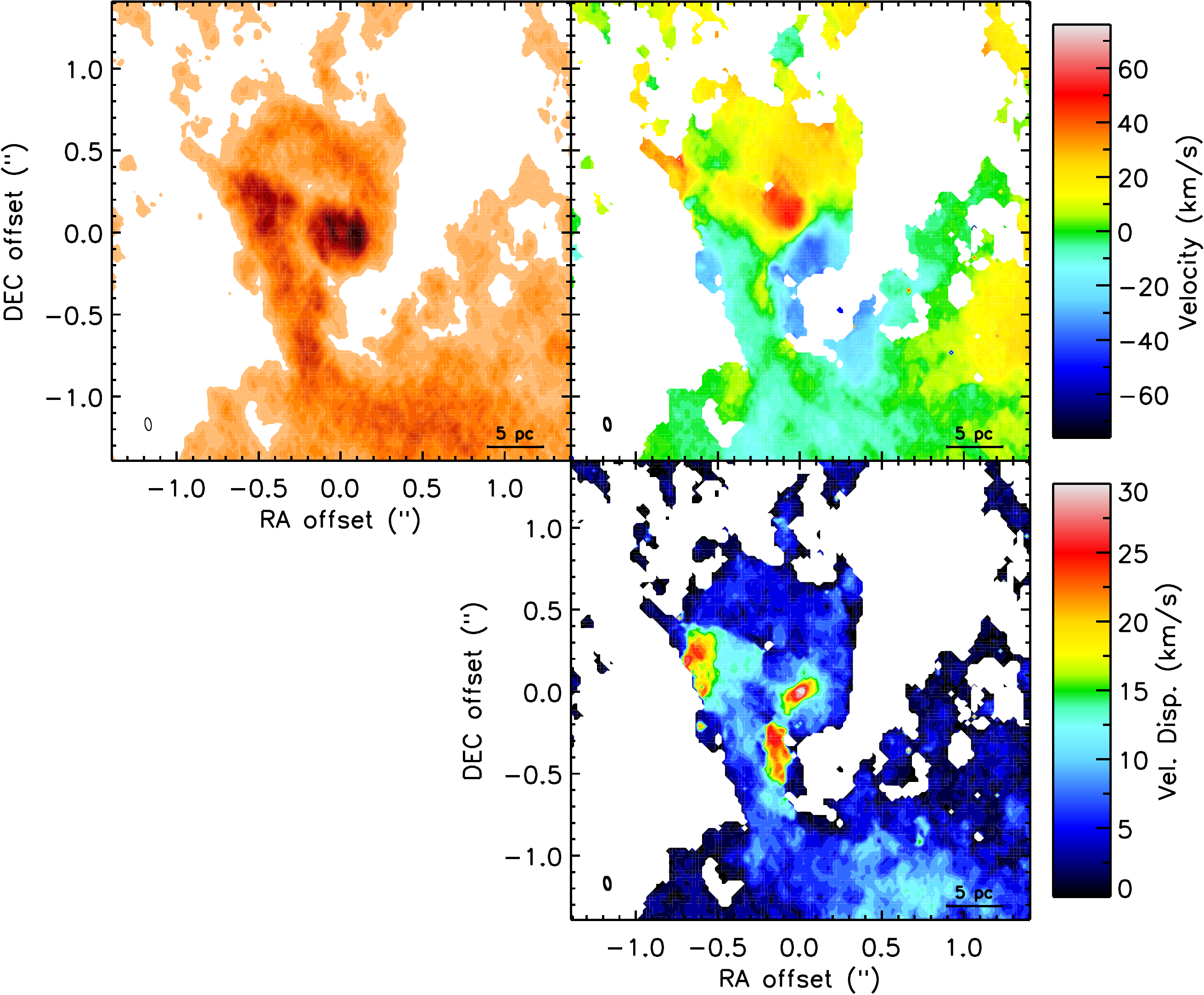}
\caption{As for Figures \ref{ALMAmoments} and \ref{ALMAkinmoments}, but for the central 2\farc8$\times$2\farc8 ($\approx$42$\times$42 pc) of NGC\,404 only. The rotating gas in the central disc of NGC\,404 can clearly be seen, along with a molecular gas one-armed spiral or arm connecting this disc to the ring present at large radii. }
\label{ALMAmoments_zoom}
 \end{center}
 \end{figure*}
 
\subsection{Line emission}
\label{lineemiss}
We clearly detect CO(2-1) line emission in NGC\,404. In our combined dataset we find an integrated line flux of $161\pm16$\,Jy\,\kms\ within the inner 8\arcsec$\,\times$\,8\arcsec ($\approx$120 pc$^2$). This error is dominated by the $\approx$10\% flux calibration uncertainties. The high-resolution dataset resolves out 74\% of this emission, which is why we only use it for kinematic modelling.
     
\cite{2015AJ....149..187T} presented a low-resolution map of CO(1-0) in this source, and combining their observed flux density with our own suggests a CO(2-1)/CO(1-0) ratio in beam temperature units of $\approx$0.6, within the range found in normal spiral galaxies \citep[e.g.][]{2009AJ....137.4670L}. To check if we are likely to be resolving out flux even in our combined dataset, we reduced the ALMA Compact Array (ACA) data available for this source from project 2017.1.00907.S. The gas in NGC\,404 is barely resolved by the ACA, and this data should thus provide a good measurement of the total flux. We derive a total CO(2-1) flux of $174\pm17$ Jy\,\kms\ from these data, in agreement with the estimate from our combined dataset. This suggests that we are not resolving out significant flux in our combined dataset, despite its extremely high spatial resolution. 
  
Zeroth (integrated intensity), first (mean light-of-sight velocity) and second (line-of-sight velocity dispersion) moment maps of the detected line emission were created from the combined datacube using a masked moment technique (see \citealt{2011arXiv1101.1499D}).  These moments are presented in Figures \ref{ALMAmoments} and \ref{ALMAkinmoments}. 
Figure \ref{ALMA_HST_over} shows moment zero contours over-plotted on \textit{Hubble Space Telescope} (\textit{HST}) images of the stellar light (in the F336W filter), and ionised gas (H$\alpha$+[\nii] emission from \textit{HST} narrow-band imaging) of NGC\,404. These \textit{HST} images were calibrated astrometrically by \cite{2017ApJ...836..237N} and the ALMA data are overlaid using their native astrometric solution based on the known position of the phase-calibrator sources.  A major-axis position-velocity diagram (PVD)  was extracted from the same data cube with a position angle of 37$^{\circ}$ {(the best-fit position angle determined for the central disc of molecular gas, see Section \ref{model}}) and a width of 5 pixels and is shown in Figure \ref{ALMA_pvd}. Coordinates in these plots are shown relative to the phase centre of the ALMA data, International Celestial Reference System (ICRS) position 01$^{\rm h}$09$^{\rm m}$27\coordsec001, +35$^{\circ}$43$^{\prime}$04\farc942.

The morphology of the molecular gas in NGC\,404 is complex. A clearly rotating central disc/torus structure is present within the central $\approx$5\,pc, at the same location as the hot-H$_2$ disc \citep{2010ApJ...714..713S}. A zoomed view of the central region of the galaxy, highlighting this disc, is shown in Figure \ref{ALMAmoments_zoom}. The position of the apparent kinematic centre of this disc coincides with the peak of the spatially-resolved nuclear radio continuum source \citep[possibly associated with a confined jet;][]{2017ApJ...845...50N}, suggesting the disc is rotating around the central accreting BH. It is the kinematics of this central material that we will concentrate on to determine the black hole mass (Section \ref{model}).  The emission outside this region is dominated by a ring like structure containing multiple highly-resolved molecular clouds and filaments, that correspond to dust features seen in absorption in the \textit{HST} images (Figure \ref{ALMA_HST_over}). The morphology and kinematics of these clouds will be discussed in detail in Section \ref{gasmorph} and Liu et al. (in prep).

 \begin{table*}
\caption{Continuum sources detected towards NGC\,404 at 237 GHz.}
\begin{center}
\begin{tabular*}{0.8\textwidth}{@{\extracolsep{\fill}}l r r r r r r}
\hline
Description & Label & RA & Dec &Ang. size &Physical size& Flux density\\
& &  (hh:mm:ss.sss) & ($^{\circ}$:$^{\prime}$:\arcsec) & & (pc$^2$) & (mJy)\\
\hline
Centre + arm &  A &01:09:26.999 & +35:43:05.02 & 1\farc55 $\times$ 0\farc9 & 23 $\times$ 13 & 0.38$\pm$0.02\\
East &  B &01:09:27.175 & +35:43:05.17 & $<$0\farc7 & $<$10 $\times$ 10 & 0.09$\pm$0.02\\
South-west & C & 01:09:26.886 & +35:43:03.93 & $<$0\farc7 & $<$10 $\times$ 10 & 0.09$\pm$0.02\\
North  &  D & 01:09:27.133  & +35:43:14.81 & $<$0\farc7 & $<$10 $\times$ 10 & 0.20$\pm$0.02\\ 
\hline
\end{tabular*}
\end{center}
\label{conttable}
\end{table*}

\subsection{Continuum emission}

\begin{figure}
\begin{center}
\includegraphics[height=7.5cm,angle=0,clip,trim=0.0cm 0.0cm 0cm 0.0cm]{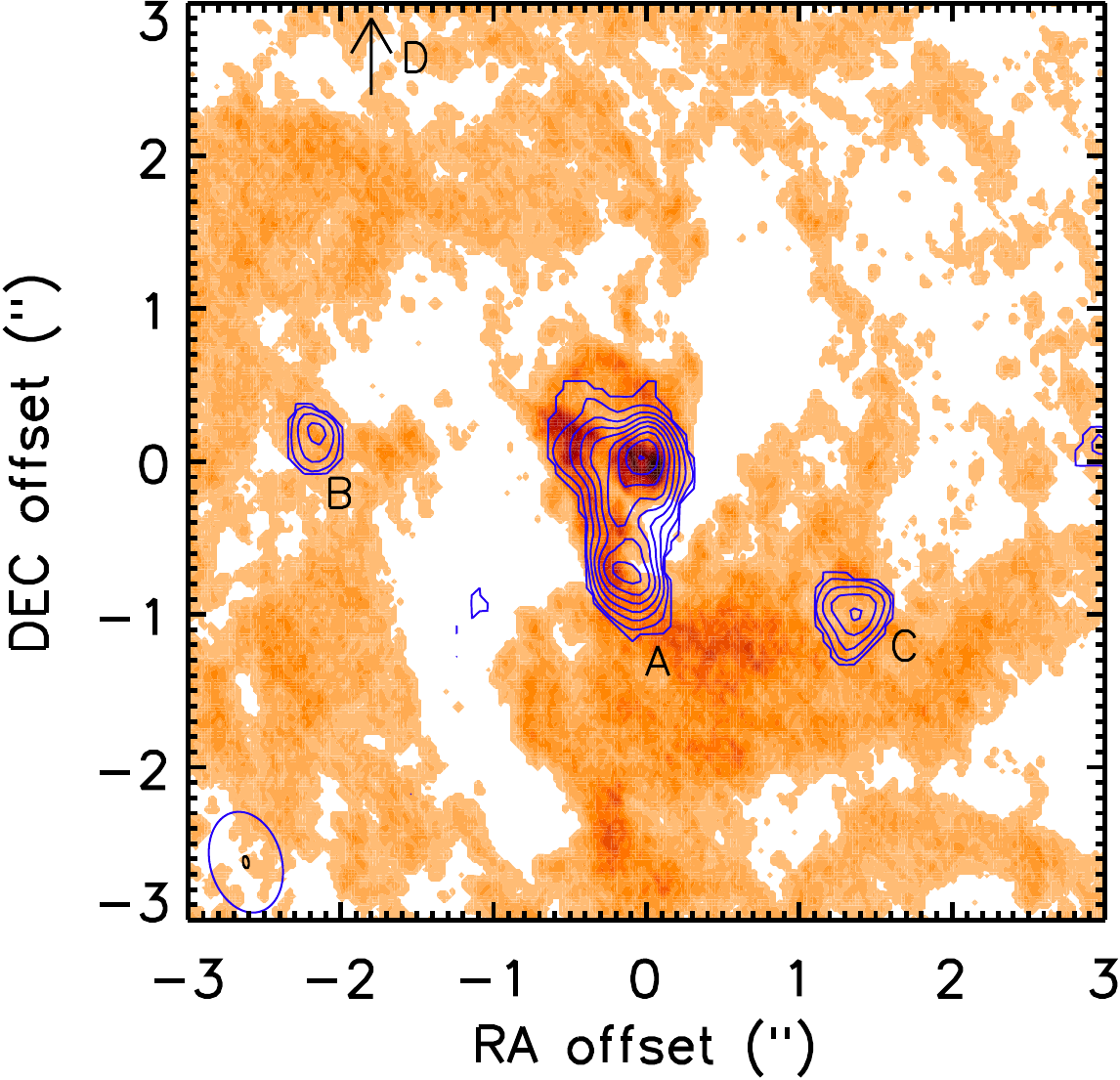}
\caption{ALMA 237 GHz continuum emission (blue contours) overlaid on an integrated intensity map of the CO(2-1) emission in NGC\,404 (orange). The continuum is imaged with a synthesised beam of 0\farc67\,$\times$\,0\farc46 (or $\approx$9.9$\times6.8$\,pc) as shown in blue in the bottom-left corner, while the 0\farc078\,$\times$\,0\farc037 (or $\approx$0.8 pc$^2$) synthesised beam of the underlying CO(2-1) map is shown in black. An extended central continuum source is detected coincident with the molecular material, along with several point-like sources further out in the molecular ring. A fourth source is detected well outside the area shown in this figure.}
\label{ALMAcont_moments}
 \end{center}
 \end{figure}

We imaged the continuum emission of NGC\,404 at a mean frequency of 237\,GHz using the line free region of all spectral windows from all observed tracks.  We used a multiscale clean algorithm, as discussed above, and additionally tapered the data in the $uv$-plane (yielding a synthesised beam of 0\farc67\,$\times$\,0\farc46 or $\approx$9.9$\times6.8$\,pc) to maximise our sensitivity to diffuse, extended emission. The resulting continuum image has an RMS noise of 17$\mu$Jy beam$^{-1}$. Continuum emission was detected from four different positions in NGC\,404. Figure \ref{ALMAcont_moments} shows the inner three sources as blue contours, over-plotted on the CO(2-1) combined moment zero map. The final source is located well to the north, outside the CO(2-1) detection region. The centroid positions and integrated fluxes of these four emission regions are tabulated in Table \ref{conttable}. 
We return to the nature of these sources in Section \ref{cont_discuss}.

\begin{figure*}
\begin{center}
\includegraphics[width=0.85\textwidth,angle=0,clip,trim=0.0cm 0.0cm 0cm 0.0cm]{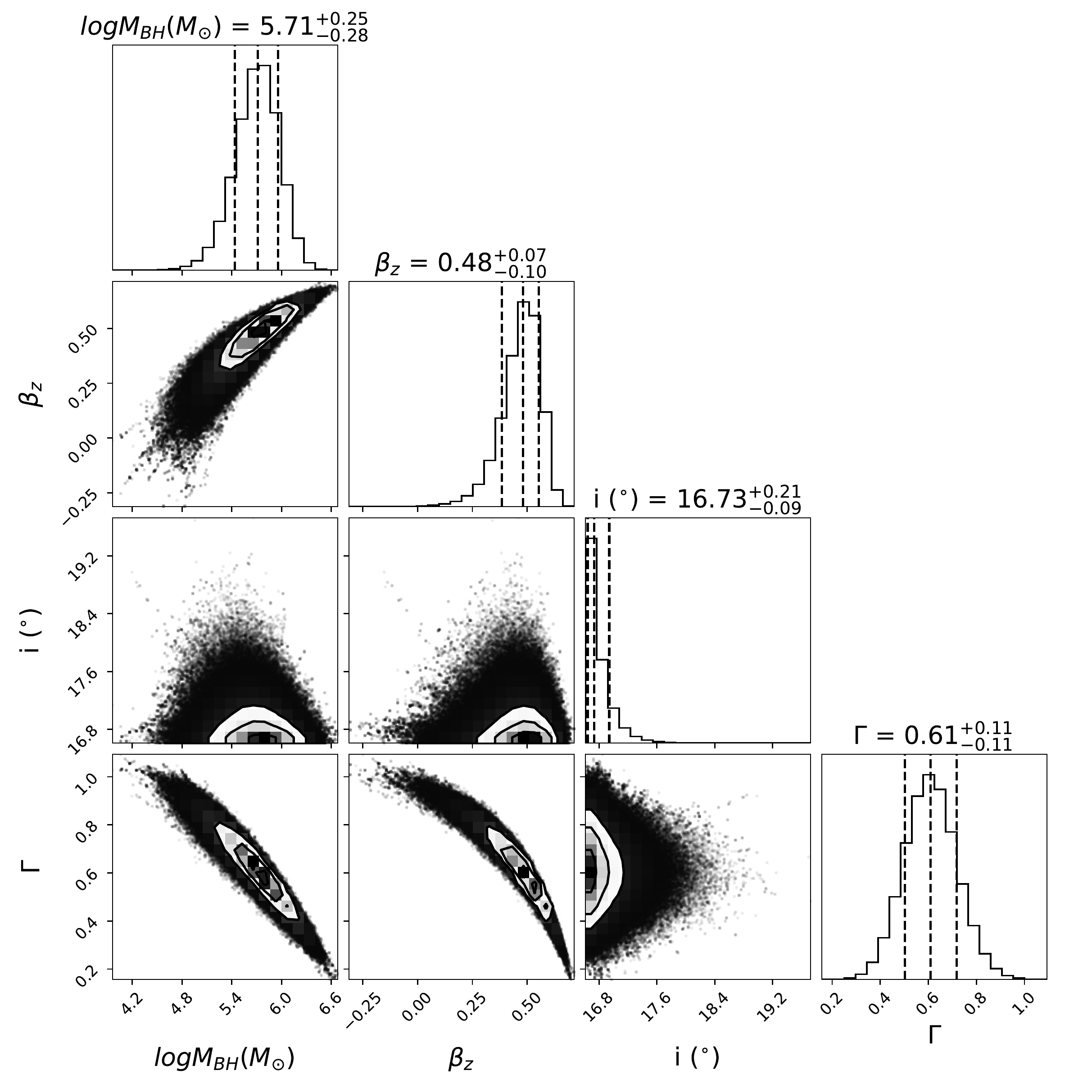}
\caption{Multi-dimensional parameter space posterior distributions calculated for our JAM stellar-dynamical model of NGC\,404. Top panels in the diagonal are marginalised one-dimensional histograms, each showing the distribution of the corresponding column parameter. The median value and 68\% (1$\sigma$) confidence levels are indicated with black dashed vertical lines. In the panels below, the three contours show the two-dimensional marginalisations of the fitted parameters at the 1$\sigma$, 2$\sigma$ and 3$\sigma$ confidence levels.}
\label{stellar_cornerplot}
 \end{center}
 \end{figure*}

\section{Stellar kinematic modelling}
\label{stellar_model}

Limits on the mass of the BH in NGC\,404 were estimated in the past \citep{2010ApJ...714..713S, 2017ApJ...836..237N} using both stellar kinematics and gaseous tracers (hot-H$_2$ lines in the near-infrared). As discussed above, while the limits from these measurements overlap, the best-fitting BH masses do not agree, and it has been posited that this is because the hot-H$_2$ gas is kinematically disturbed. However, the previous kinematic measurements included a model of the stellar mass in this object, but did not take into account the mass of the molecular gas component.  As we show in Figure \ref{ALMAmoments}, molecular gas is present in significant quantities in the vicinity of the BH in this object, and could thus affect measurements of the BH mass.  If the gas were concentrated primarily around the BH then clearly this would lead to an overestimate of the BH mass \cite[e.g.][]{2019ApJ...885L..21M}, as the molecular gas mass would be wrongly ascribed to the central point mass. If, on the other hand, the gas were distributed more widely, but not following the stellar mass distribution, the stellar kinematic models would then be forced to assign a higher mass-to-light ratio to the stellar component to account for this unseen mass. In some cases this could cause the models to \textit{under}-estimate the BH mass.  

To assess the impact of the molecular material on the stellar kinematic BH mass measurement in NGC\,404 we re-run the stellar kinematic modelling procedure of \cite{2017ApJ...836..237N}, now taking into account the gravitational potential of the molecular gas. To do this we create a multi-Gaussian expansion (MGE; \citealt{Emsellem:1994p723, 2002MNRAS.333..400C}) model of the molecular material, using the python \textsc{MgeFit} package to fit the molecular gas surface density map\footnote{Available from \url{https://pypi.org/project/mgefit/}}.  The resulting MGE is tabulated in Table \ref{mgetable}. This model does not include the flocculent sub-structure of this system, but allows us to quantify the contribution to the potential of an ideal axisymmetric version of this observed molecular gas disc. We have assumed that the mass of any \hi\ present around the nucleus of NGC\,404 is negligible. This seems reasonable, given that the molecular gas in the centre of this object has a much higher surface density ($\sim$10,000 \msun\ pc$^{-2}$) than usually found for \hi\ (that typically saturates at $\sim10$\,\msun\,pc$^2$), and that \hi\ observations of this object reveal a central hole \citep{2004AJ....128...89D}. {The mass of hot-H$_2$ emitting molecular gas in the nucleus of this galaxy is similarly not included, but as it has a total mass of $<1$\msun\ \citep{2010ApJ...714..713S} it is dynamically unimportant}. We are thus able to estimate the total gas mass from the molecular gas data alone, using the CO(2-1)/CO(1-0) ratio as calculated above and a Galactic CO-to-H$_2$ conversion factor of 3$\times$10$^{20}$ cm$^{-2}$ (K \kms)$^{-1}$ \citep{1986ApJ...309..326D}. We use a Galactic CO-to-H$_2$ conversion factor for NGC\,404, as despite its low stellar mass it has approximately solar metallicity \citep{2013ApJ...772L..23B}. We discuss further the impact of this assumption in Section \ref{uncerts}.

This model of the gas potential was then included with the stellar mass model of this system, along with an unknown BH mass. We use here the MGE model of the stellar mass distribution from \cite{2017ApJ...836..237N}, constructed from \textit{HST} images of the nucleus, using mass-to-light ratios calculated on a pixel-by-pixel basis from multi-band colours (that should also correct for stellar population gradients and dust obscuration).  We note that the lowest inclination allowed by this model of the stellar potential is 16\fdeg5. Using this combined mass model we performed a fit to the observed stellar kinematics using the Jeans anisotropic modelling (JAM) code of \cite{2008MNRAS.390...71C}, relying on an axisymmetric solution of the Jeans equations incorporating orbital anisotropy. Full details of the fitting procedure can be found in \cite{2017ApJ...836..237N} {and \citet{2019ApJ...872..104N}}.  The free parameters of this fit are the BH mass,  orbital anisotropy, inclination of the system and stellar mass normalisation factor $\Gamma$ (where $\Gamma=1$ corresponds to the mass derived from stellar population fitting in \citealt{2017ApJ...836..237N} assuming a \citealt{2003PASP..115..763C} initial mass function; IMF). These parameters were given flat priors (or flat in log space for the BH mass) and were allowed to vary within reasonable ranges (listed in Table \ref{jamtable}).

\begin{table}
\caption{MGE fit to the molecular gas distribution in NGC\,404.}
\centering
\begin{tabular}{rrr} 
\hline 
$\Sigma_{H_2}$  & $\sigma$ &  $q$\\
(\msun\ pc$^{-2}$) & (arcsec) & \\
\hline
      13304.7  &     0.082      & 1.00\\
      168.6    &   0.091      & 1.00\\
      1409.4     &  0.380     &  1.00\\
      975.9     &  2.187   &    1.00\\
      38.2     &  4.492 &      1.00\\
\hline
\end{tabular}
\parbox[t]{0.4\textwidth}{ \textit{Notes:} For each fitted component of the MGE, Column 1 lists its central molecular gas surface density, Column 2 its standard deviation (width) and Column 3 its axis ratio. All quantities are intrinsic (i.e. deconvolved). }  
\label{mgetable}
\end{table}

\begin{table}
\caption{JAM model best-fitting parameters and statistical uncertainties.}
\centering
\begin{tabular}{lcrrr} 
\hline     
Parameters          &    Range    &Best Fit&$1\sigma$ Error &$3\sigma$ Error\\
                    &                   &        &   (68\%)       &     (99\%)  \\  
  (1)               &        (2)        &   (3)  &      (4)    &      (5)      \\  
\hline
log$_{10}$($M_{\rm BH}$/\msun)&  2~$\rightarrow$~7  &  5.71 &$-$0.28, +0.25&$-$0.91, +0.59\\
$\beta_z$           &$-1 \rightarrow +1$ &  0.48 &$-$0.10, +0.07&$-$0.32, +0.18\\
$i$ ($^{\circ}$)    &16.5~$\rightarrow$~90.0& 16.73 &$-$0.09, +0.21&$-$0.13, +0.60\\
$\Gamma$            &0.1~$\rightarrow$~1.7& 0.61  &$-$0.11, +0.11&$-$0.30, +0.30\\
\hline
\end{tabular}
\parbox[t]{0.5\textwidth}{ \textit{Notes:} Columns 1 and 2 list the fitted model parameters, and their search ranges. The priors are uniform in linear space (except that of $M_{\rm BH}$ which is uniform in logarithmic space). Columns 3, 4 and 5 show the best-fitting value of each model parameter and its uncertainties at 1$\sigma$ and 3$\sigma$ confidence levels.}  
\label{jamtable}
\end{table}

 \begin{figure}
\begin{center}
\includegraphics[width=0.5\textwidth,angle=0,clip,trim=0.0cm 0.0cm 0cm 0.0cm]{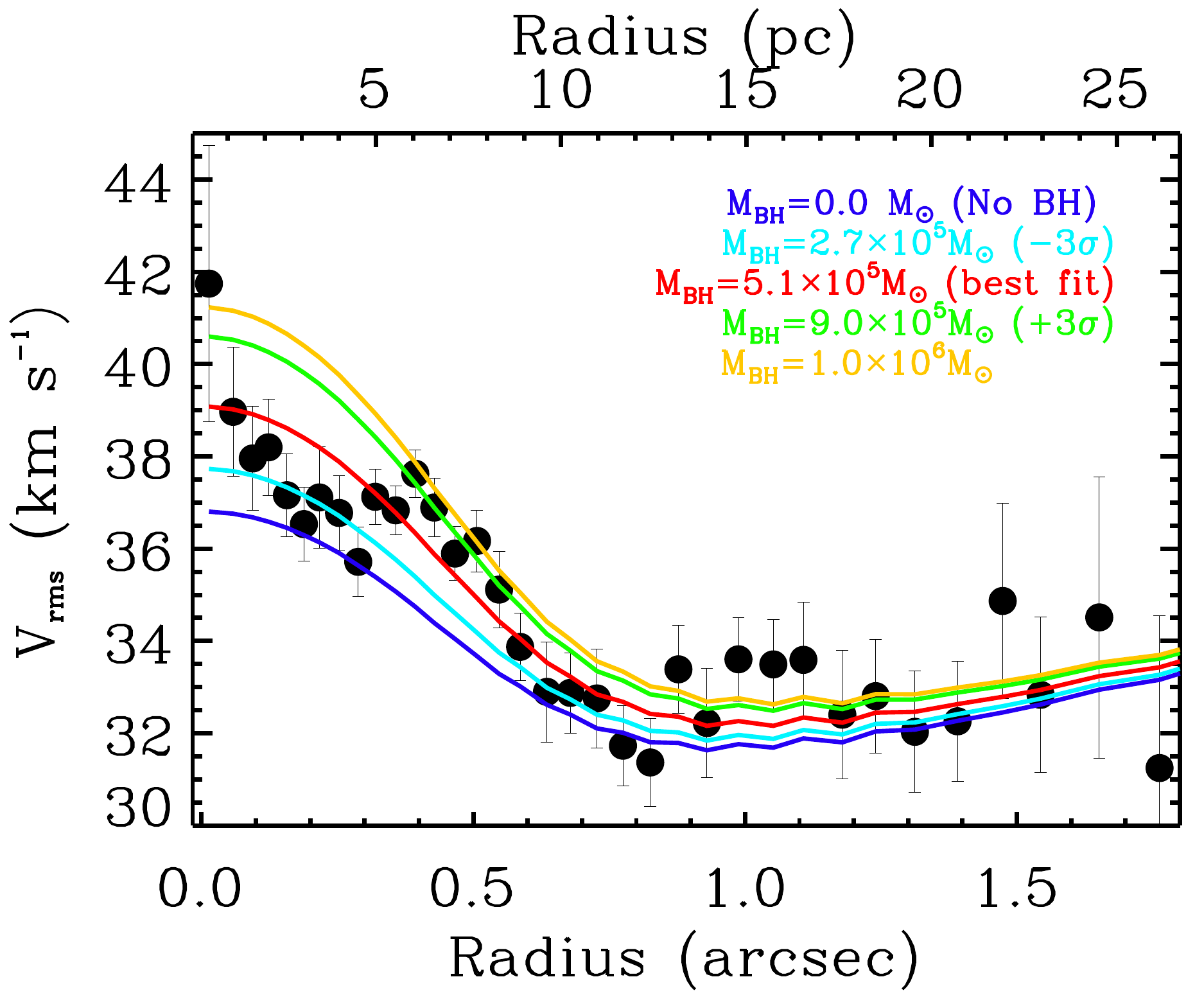}
\caption{One-dimensional radial profile of the RMS velocity ${V}_{\mathrm{rms}}$ vs. our JAM prediction for axisymmetric mass models with different BH masses as listed in the legend. All models are fixed to the best-fitting anisotropy parameter ($\beta_z=0.5$), mass scaling factor ($\Gamma=0.61$) and inclination angle ($i=16.7^{\circ}$). The data are binned radially in bins of width $0.05''$ (or 0.7 pc). Our best-fitting model clearly favours an IMBH of mass (2.7--9.0)$\times10^5$\,\msun.}
\label{stellar1drms}
 \end{center}
 \end{figure}

Figure \ref{stellar_cornerplot} shows a visualisation of the multi-dimensional parameter space explored by our fit to the stellar kinematic data. In the top panel of each column, a one-dimensional histogram shows the marginalised posterior distribution of that given parameter, with the median and 68\% (1$\sigma$) confidence interval indicated by vertical black dashed lines. In the panels below, the greyscale regions show the two-dimensional marginalisations of the fitted parameters, based on 3 million iterations. While the inclination of the system is at the lower bound allowed by the MGE model, the BH mass and stellar mass normalisation factor $\Gamma$ are well constrained (see Table \ref{jamtable}).  We find that the best-fitting BH mass of NGC\,404 is 5.1$^{+3.9}_{-2.4}\times$10$^5$\,\msun\ and the stellar mass normalisation factor is $\Gamma=0.61\pm0.11$ (at the 68\% confidence level). This best-fitting JAM model is the median of the full posterior distribution, that has a reduced-chi$^2$ $\chi^2_{\rm red}=1.23$ over 920 degrees of freedom \citep{2017ApJ...836..237N}. 

To better illustrate that our new JAM-stellar dynamical model fits the data with the aid of the additional molecular gas component, Figure \ref{stellar1drms} shows the one-dimensional radial profile of the RMS observed velocity (${V}_{\mathrm{rms}}$) (as presented in \protect \citealt{2017ApJ...836..237N}) versus our JAM predictions for axisymmetric mass models with different BH masses. 
While none of the models provide a perfect match to the data, BHs with ${M}_{\mathrm{BH}}\ltsimeq$10$^5$\,\msun\  and $\gtsimeq$10$^6$\,\msun\ are unable to fit the data from the inner parts of NGC\,404 (R$\ltsimeq$0\farc5 or 7.4\,pc). 

\begin{figure}
\begin{center}
\includegraphics[width=0.45\textwidth,angle=0,clip,trim=0.0cm 0.0cm 0cm 0.0cm]{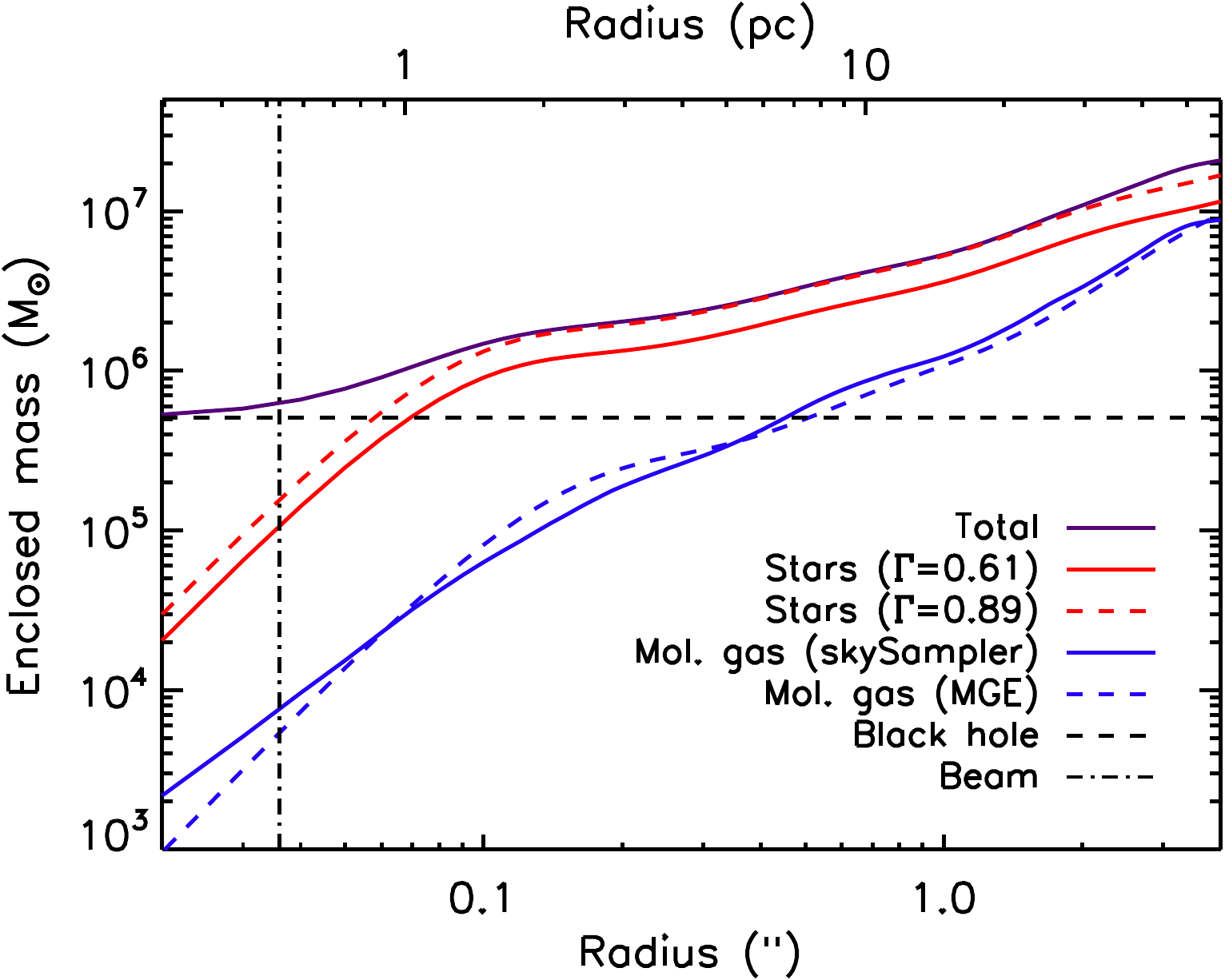}
\caption{Enclosed mass of NGC\,404 as a function of radius from our kinematic modelling (purple solid line). The contribution to the total mass from stars (with our best fitting $\Gamma=0.61$) is plotted as the red solid line, that from molecular gas as the blue lines (where the solid blue line shows the estimate from our \textsc{skySampler} model, and the dashed blue line our MGE model estimate) and that from the black hole as the black dashed line. The red dashed line shows the stellar contribution with $\Gamma=0.89$ (as found by \protect \citealt{2017ApJ...836..237N}), that would provide a good fit to the total mass enclosed if one did not know about the presence of molecular gas (that contributes significantly to the enclosed mass, especially at larger radii).}
\label{N404_massplot}
 \end{center}
 \end{figure}

This best-fitting BH mass is significantly higher than that previously derived from the same data, while the stellar mass normalisation factor is significantly lower (\citealt{2017ApJ...836..237N} found a 3$\sigma$ upper limit on the black hole mass of $1.5\times10^5$\,\msun\ and $\Gamma=0.89^{+0.06}_{-0.05}$).  {This difference is not driven by the presence of molecular gas around the BH itself, but rather by the increasing contribution of molecular gas to the potential at larger radii ($>2$\,pc).} We show this graphically in Figure \ref{N404_massplot}, plotting the cumulative mass within NGC\,404 as a function of radius. The stellar and molecular gas contributions to the total mass budget are shown as red and blue solid lines, respectively. The stellar contribution has been estimated from our best-fitting JAM model parameters. The molecular gas contribution has been calculated in two independent ways, the first from our MGE model of the gas surface density (blue dashed line), the second from the clean components of the ALMA data (as discussed in Section \ref{molgasdist_mod}; blue solid line). These estimates differ slightly but are very similar in shape and normalization. While the molecular material does not dominate the potential at any radius, it does provide a significant contribution towards the outer edge of the central molecular gas disc/torus (contributing 16\% of the luminous mass within 0\farc25). At around 1\arcsec\ (the edge of the field of view of the Gemini data used to measure the stellar kinematics), this contribution grows to $\approx$35\%. The median contribution of the molecular gas within this field of view is 26\%. The contribution of the molecular gas to the gravitational potential peaks at 80\% at $\approx$4\arcsec\ in the large-scale molecular ring, before decreasing outside this.  As shown by the red dashed line, if one were to ignore the presence of this molecular gas, then $\Gamma=0.89$ for the stellar component (as found by \citealt{2017ApJ...836..237N}) would match well the total mass profile. {When one includes the potential of the molecular gas a much lower stellar mass normalisation factor ($\Gamma=0.61$) is required. This lower stellar mass normalisation decreases the amount of stellar mass present in the very central regions around the BH in our model, and thus a larger BH is required to fit the kinematics.}

\section{Molecular gas kinematic modelling}
\label{model}

In the preceeding Section we used our extremely high-resolution map of the molecular gas in NGC\,404 to include its contribution to the gravitational potential in our stellar kinematic modelling. This allowed us to obtain a new estimate of the BH mass in this low-mass galaxy. However, our ALMA observations can reveal more about the BH in this system. High-resolution molecular gas observations have been increasingly used in recent years to directly constrain BH masses \citep{2013Natur.494..328D,2015ApJ...806...39O,2016ApJ...822L..28B,2016ApJ...823...51B,2017MNRAS.468.4663O,2017MNRAS.468.4675D,2018MNRAS.473.3818D,2019MNRAS.485.4359S,2019MNRAS.490..319N,2019A&A...623A..79C,2019ApJ...881...10B,2019ApJ...883..193N,2019arXiv190203813N}. In this Section, we employ this technique to obtain an independent measurement of the mass of the IMBH in NGC\,404.

To do this we concentrate on the molecular gas disc/torus within a radius of 0\farc25 (or $\approx$3.7\,pc) from the centre of NGC\,404. We create a dynamical model of the molecular gas in this disc, including the gravitational influence of the stars and molecular gas along with that of the black hole. We use a forward modelling approach to compare the observed datacube with our model, and thus derive the best-fitting BH mass (and molecular gas disc parameters). 

To achieve this we use the Kinematic Molecular Simulation (\textsc{KinMS}\footnote{https://github.com/TimothyADavis/KinMS}) mm-wave observation simulation tool of \cite{2013MNRAS.429..534D}, coupled to the Markov Chain Monte Carlo (MCMC) code \textsc{KinMS\_mcmc\footnote{https://github.com/TimothyADavis/KinMS\_MCMC}}. This tool allows input guesses for the true gas distribution and kinematics and, assuming the gas is in circular rotation, produces a simulated data cube that can be compared to the observed data cube (taking into account the effects of beam-smearing, spatial and velocity binning, disc thickness, gas velocity dispersion, etc.).
This tool has been applied to estimate BH masses in various other works \citep{2017MNRAS.468.4663O,2017MNRAS.468.4675D,2018MNRAS.473.3818D,2019MNRAS.485.4359S,2019MNRAS.490..319N,2019arXiv190203813N}, each of which has iteratively improved the technique. We highlight the important aspects of this fitting procedure below.

\subsection{Gas distribution}
\label{molgasdist_mod}

One of the inputs of \textsc{KinMS} is an arbitrarily parameterised gas surface brightness distribution. In most of the previous papers using this technique, an axisymmetric molecular gas distribution was assumed and fit to the datacube. As the molecular gas distribution in NGC\,404 is clearly not axisymmetric we here utilise a different approach, making a model of the gas distribution based on the clean components generated when cleaning our observational data. This allows us to remove the uncertainty (and possible issues with model mismatch) caused by fitting the morphology of the gas distribution, as it is reproduced well by construction. We use the publicly available \textsc{skySampler}\footnote{https://github.com/Mark-D-Smith/KinMS-skySampler} python code of  \cite{2019MNRAS.485.4359S}, that samples the observed clean components with an arbitrary number of point sources ($10^6$ in this case). These point sources have a uniform brightness, so more are generated in locations with bright molecular gas emission. These point sources can be de-projected at each step of our MCMC sampling algorithm to accurately reproduce the gas distribution observed at any proposed position angle and inclination. Each point source can then be assigned a velocity based upon our kinematic model, described below.

\subsection{Mass model}
\label{stellerdist}

The molecular gas in galaxies moves in response to the total gravitational potential of the galaxy, that includes the mass of both luminous (stars, gas) and dark (IMBH, dark matter) components. 
As above we use a mass model of NGC\,404 to account for the contribution of luminous matter, and infer the mass of any dark object present. We include a BH in the model as a simple point mass at the kinematic centre of the galaxy. 

To take into account the distribution of the stars in NGC\,404, we again use the MGE model of the stellar mass distribution from \cite{2017ApJ...836..237N}, as above. We fix the inclination used to de-project this mass distribution to that found in Section \ref{stellar_model}, although we note that the MGE model at the relevant radii is approximately spherical, so our results are insensitive to this choice.  While in principle this model should already well describe the stellar mass, as before we include a mass normalisation factor ($\Gamma$) that can account for any systematic change in e.g the assumed IMF, star-formation and chemical-enrichment histories of the stellar populations. In Section \ref{stellar_model} we found $\Gamma$=$0.61\pm0.11$ provides the best fit to the observed stellar kinematics. This mass normalisation parameter is free in our molecular gas kinematic model, but we use this earlier determination as a prior to constrain our fit. 

We note that this model does not formally include dark matter, as it is expected to be dynamically unimportant in the inner few parsecs of this galaxy, given its relatively high stellar surface density \citep[e.g.][]{2016ApJ...827L..19L}. If dark matter were significant, its effect would simply be subsumed into the mass normalisation factor (unless it has a density distribution markedly different from that of the stars).

In most massive galaxies, the stellar mass completely dominates. In NGC\,404, however, we have shown above (Fig. \ref{N404_massplot}) that the mass of the molecular material also matters, so we include it when calculating the circular velocity of the system. The molecular gas gravitational potential is calculated from a symmetrised version of our \textsc{skySampler} cloudlets, assuming they lie in a thin but warped disc {(we discuss the impact of this thin disc assumption in Section \ref{uncerts})}. The potential arising from this gas distribution is calculated using the Interactive Data Language (IDL) \textsc{POTENTIAL\_NEWTON} routine\footnote{From Jeremy Bailin's IDL Utilities\\ \url{http://simulated-galaxies.ua.edu/jbiu/}}, taking into account its full three-dimensional distribution, and assuming each cloudlet has an equal fraction of the total gas mass.  
Within a radius of $\approx$0\farc25 (3.7\,pc) from the centre of NGC\,404 (the region included in our kinematic model) the total molecular gas mass is (2.5$\pm$0.3)$\times$10$^5$ \msun\ (where the uncertainty in this quantity is dominated by the flux calibration). We note that if instead of using this cloudlet based method we had used the MGE model of the molecular gas potential derived in Section\,\ref{stellar_model}, our results would not have changed (see also Fig. \ref{N404_massplot}).

\subsection{Gas kinematics}

As Figure \ref{ALMAkinmoments} shows, the position angle and inclination of the gas disc in NGC\,404 are clearly not radially constant (as also discussed by other authors, e.g. \citealt{2004AJ....128...89D}), although they do seem approximately fixed inside the very compact molecular gas disc/torus. 
We thus fit the data in the central $\approx$3\,pc of this source assuming a thin disc with radially constant but free inclination/position angle. Outside of this radius, where the molecular gas is concentrated in an arm-like feature, the gas distribution warps, continuously changing both its inclination and position angle (PA) with radius, before become flat again in the molecular \mbox{(pseudo-)ring} ($R\gtsimeq$15 pc). Fitting our large-scale data following the procedure outlined below yields $i=$9\fdeg3 and PA=1$^{\circ}$ for the outer parts of this object, in agreement with the values derived by \cite{2004AJ....128...89D} for the (even larger scale) \hi\ ring. The simple model that we adopt thus includes the following features (listed as a function of radius); a flat disc of radially constant but free inclination and PA within the inner 3\,pc, a warped component between 3\,pc and 15\,pc (where the inclination and PA change linearly with radius), and a further un-warped disc at $>$15\,pc with parameters fixed to those found above ($i=$9\fdeg3 and PA=1$^{\circ}$).
We note that in order to estimate the BH mass in this system we confine our kinematic model to galactocentric radii $\leq$0\farc25 ($\approx$3.7\,pc), and as such the warped material has very little effect on the fit. Using a simple single un-warped disc model does not result in a significant change in any of the derived parameters, although it does slightly increase their uncertainties. 

We also include a parameter for the internal velocity dispersion of the gas, that is assumed to be spatially constant. 
We note that our treatment of the velocity dispersion is not fully self-consistent (i.e. we do not solve Jeans' equations but simply add a Gaussian scatter to our velocities). Given the low velocity dispersion in molecular gas clouds this is not expected to significantly influence our results, but we discuss this issue further in Section \ref{uncerts}.

In this work we assume that the molecular gas is in circular motion, and hence that the gas rotation velocity varies only radially. Warps like that discussed above are likely to induce some non-circular motions, that could affect our dynamical measurements. We do not expect this to be important in the (apparently) un-warped central molecular gas torus/disc around the BH, but we can search for such signatures using the residuals around our best-fitting model (see Section \ref{uncerts}).

 \begin{figure*}
\begin{center}
\includegraphics[width=0.9\textwidth,angle=0,clip,trim=0.0cm 0.0cm 0cm 0.0cm]{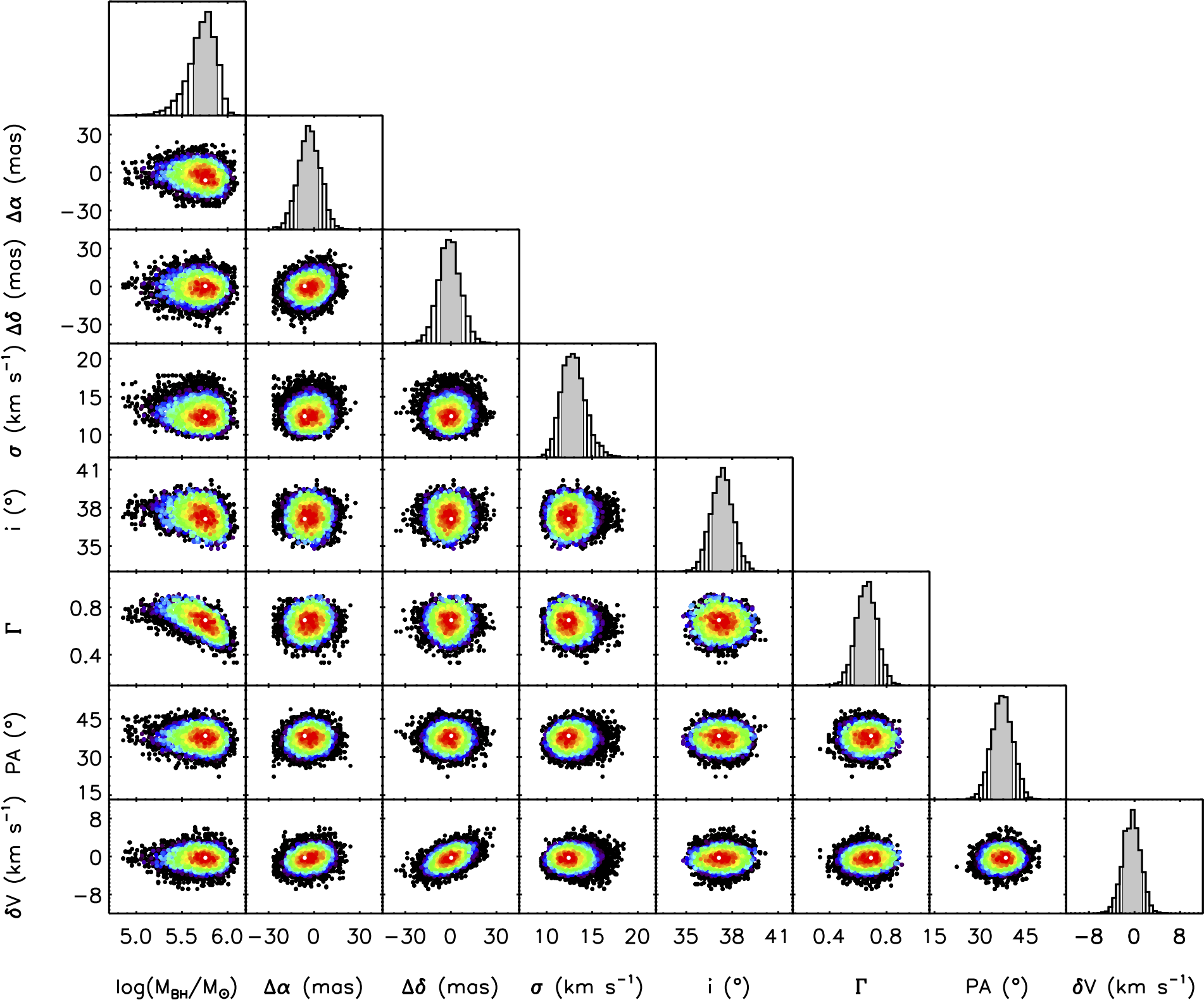}
\caption{Visualisation of the multi-dimensional parameter space explored by our molecular gas kinematic fit to the observed data of NGC\,404 within galactocentric radii $\leq$0\farc25 ($\approx$3.7\,pc). In the top panel of each column a one-dimensional histogram shows the marginalised posterior distribution of that given parameter, with the 68\% (1$\sigma$) confidence interval shaded in pale grey. In the panels below, the coloured regions show the two-dimensional marginalisations of those fitted parameters. 
Each point is a realisation of our model, colour-coded to show the relative log-likelihood of that realisation, with white points the most likely. Black points have $\Delta\chi^2$ values outside the 99\% confidence interval. See Table \ref{fittable} for a quantitative description of the likelihoods of all fitting parameters.}
\label{triangleplot}
 \end{center}
 \end{figure*}

\subsection{Bayesian analysis}
\label{fitting}

As mentioned above, we use a Bayesian analysis technique to identify the best model and estimate uncertainties. This allows us to obtain samples drawn from the posterior distributions of each model parameter. Each parameter has a prior, that we typically set as a box-car over a reasonable parameter range (an assumption of maximal ignorance). 
The prior on the black hole mass is flat in log-space, with the mass allowed to vary between log$_{10}(\frac{M_{\rm BH}}{\mathrm{M}_{\odot}})$~=~4.0 and 6.9. For the mass normalisation factor ($\Gamma$) we use a Gaussian prior based on the measurement from Section \ref{stellar_model}. We constrain the inclination angle with a Gaussian prior around the best fit found in previous work, $i$=37$\pm3^{\circ}$ \citep{2017ApJ...836..237N}. The kinematic centre of the galaxy is constrained to lie within $\pm1$\arcsec\ of the optical galaxy centre position. The systemic velocity is allowed to vary by $\pm$20 \kms\ from that found by optical analyses.  
Details of all the priors are listed in Table \ref{fittable}.

Our MCMC procedure generates a model datacube and compares it to our observed data using a simple log-likelihood ($\mathcal{L}$):
\begin{equation}
\mathcal{L}(a)=\frac{-\chi^2}{2} + \ln \left(\prod_{i=0}^{n} P(a_i)\right),
\end{equation}
where $\chi^2$ is the classical $\chi^2$ statistic, $P(a_i)$ is the prior probability of free parameter $a_i$, and $n$ is the number of free-parameters in the fit.

As discussed in detail in \cite{2019MNRAS.485.4359S}, because our ALMA data are noisy, the $\chi^2$ statistic has an additional uncertainty associated with it, following the chi-squared distribution \citep{2010arXiv1009.2755A}. This distribution has a variance of $2(N-I)$, where $N$ is the number of constraints and $I$ the number of inferred parameters. For our data $N$, the number of unmasked pixels (see Section \ref{lineemiss}), is very large ($N$=12,112), so the variance becomes $\approx2N$.
Systematic effects can produce variations of $\chi^2$ of the order of its variance \citep{2009MNRAS.398.1117V}, and ignoring this effect yields unrealistically small uncertainties. To mitigate this effect \cite{2009MNRAS.398.1117V} proposed to increase the $1\sigma$ confidence interval to $\Delta\chi^2=\sqrt{2N}$. To achieve the same effect within our Bayesian MCMC approach we need to scale the log-likelihood, as done by \cite{2017MNRAS.464.4789M}. This is achieved here by increasing the input errors (i.e. the assumed noise in the cube) by $(2N)^{1/4}$ (in our case, with $N$=12,112, this equates to increasing the RMS by a factor of 12.47). This approach appears to yield physically credible formal uncertainties in the inferred parameters, whereas otherwise these uncertainties are unrealistically small. This correction does not account for the correlations between spaxels induced by the beam of the interferometer. These correlations can be accounted for using a covariance matrix (as in e.g. \citealt{2017MNRAS.468.4675D}), but this is computationally intensive and the effect is small compared to the variance correction applied above. Hence we neglect it here. 

We utilise a MCMC method with Gibbs sampling and adaptive stepping to explore the parameter space. The algorithm runs until convergence is achieved, and then the best chain is run (with a fixed step size) for 300,000 steps (with a 10\% burn-in) to produce our final posterior probability distribution. For each model parameter the probability surfaces are then marginalised over to produce a best-fitting value (the median of the marginalised posterior samples) and associated 68\% and 99\% confidence levels (CLs). Figure \ref{triangleplot} shows the one- and two-dimensional marginalisations of the free parameters included in the fit. Quantitative descriptions of the likelihoods of all parameters are presented in Table \ref{fittable}. Most of our parameters are independent of each other, but we do find some degeneracies. As expected, the IMBH mass is degenerate with the mass-normalisation factor $\Gamma$ and the inclination (as is always the case in any BH mass fit). The nuisance parameters (the kinematic centre of the galaxy within the cube) are also slightly correlated. Despite these degeneracies almost all these parameters remain constrained by these molecular gas data alone, and any additional scatter introduced is included when marginalising to obtain our final uncertainties.  

\begin{figure*}
\begin{center}
\includegraphics[height=6cm,angle=0,clip,trim=0.0cm 0.0cm 1.9cm 0.0cm]{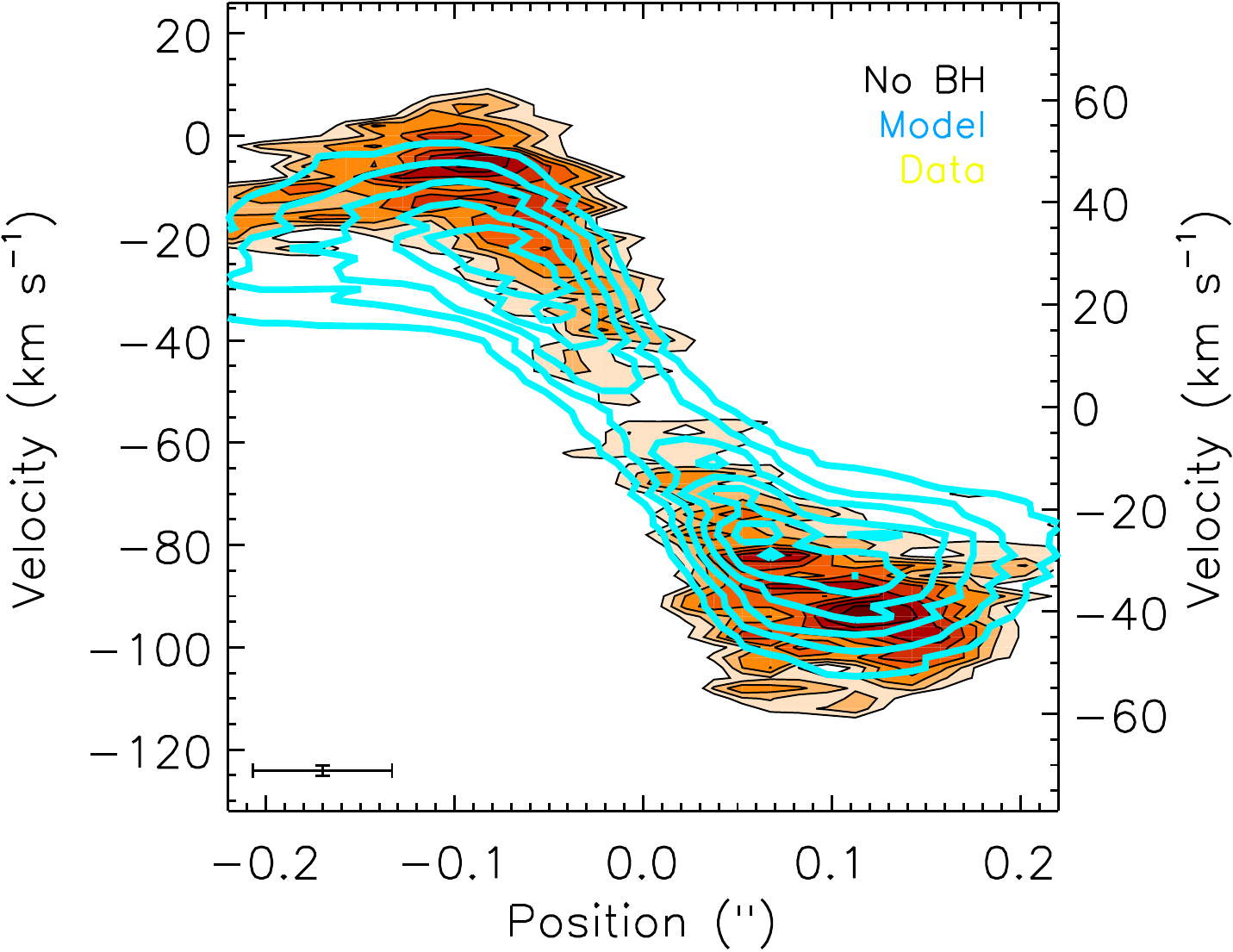}\includegraphics[height=6cm,angle=0,clip,trim=2.5cm 0.0cm 1.9cm 0.0cm]{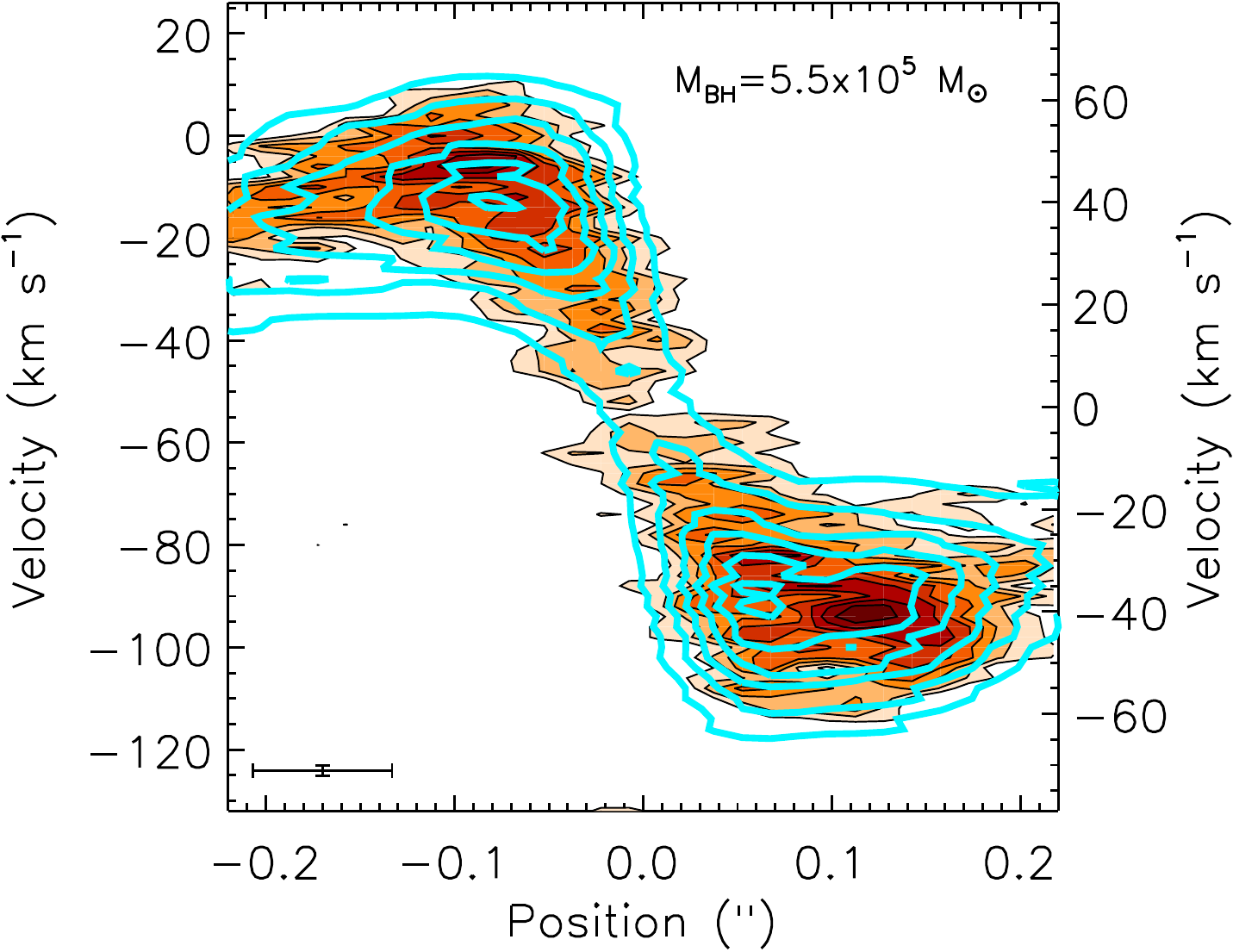}\includegraphics[height=6cm,angle=0,clip,trim=2.5cm 0.0cm 0cm 0.0cm]{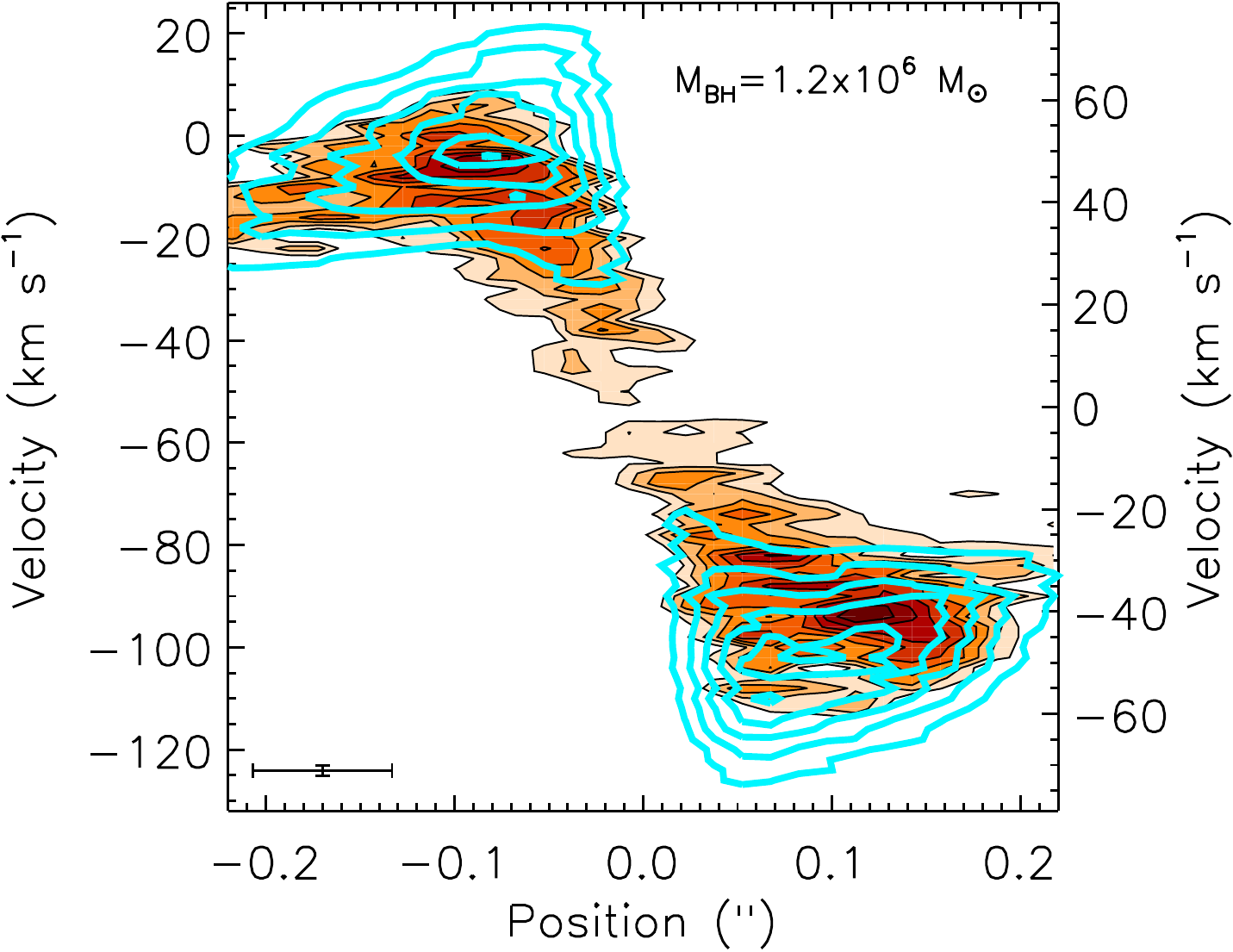}
\caption{Position-velocity diagram extracted from the observed NGC\,404 datacube (orange), overplotted with position-velocity data taken from our \textsc{KinMS} models (blue contours). Our spatial and velocity resolutions are shown as an error bar in the bottom-left corner of each plot. The central panel shows our best-fitting model with a 5.5$\times10^5$\,\msun\ IMBH, while the left and right panels show the same model with no BH, and with a SMBH ($M_{\rm BH}=1.2\times10^6$\,\msun), respectively. Clearly our best-fitting model provides a much better fit to the data.}
\label{modelpvds}
 \end{center}
 \end{figure*}

We clearly detect the presence of a dark object in the centre of NGC\,404, with a mass of 5.5$^{+4.1}_{-3.8}\times$10$^5$ \msun\ (at the 99\% CL). Our best-fitting model provides a good fit to the data, as shown by Figures \ref{modelpvds} and  \ref{vresid} that respectively show position-velocity diagrams and velocity fields extracted from our best-fitting model and compared with the observed data. As shown by Figure \ref{modelpvds}, a model without an IMBH cannot provide a good fit to our data, nor can a BH of mass $\gtsimeq10^6$\,\msun. Figure \ref{vresid} shows that the best-fitting model reproduces the observed velocity field well, leaving only low-level velocity residuals that are not spatially correlated.

 \begin{table*}
\caption{Best-fitting model parameters and statistical uncertainties.}
\begin{center}
\begin{tabular*}{0.8\textwidth}{@{\extracolsep{\fill}}l r c r r r r}
\hline
Parameter & \multicolumn{3}{c}{Prior} & Best fit & 1$\sigma$ error (68\%) & 3$\sigma$ error (99\%)\\
(1) &  \multicolumn{3}{c}{(2)} & (3) & (4) & (5)\\
\hline
Galaxy parameters:&&&&&& \\\hline
IMBH mass (log$_{10}$  M$_{\rm BH}$/M$_{\odot}$) &   4.0& $\xrightarrow[]{\rm uniform}$ &  6.9 &     5.74 &$-$ 0.07, +  0.11  & $-$  0.51, +  0.24 \\
Mass normalisation factor  ($\Gamma$) &   \multicolumn{3}{c}{$N$($\mu$=0.61, $\sigma$=0.11)} &     0.66 & $-$ 0.07, + 0.05 & $-$ 0.19, + 0.16\\
Inclination  ($^\circ$) &   \multicolumn{3}{c}{$N$($\mu$=37, $\sigma$=3)} &    37.1 & $-$ 0.46, + 0.57 & $-$ 1.76, + 1.61\\
Position angle  ($^\circ$) & 10& $\xrightarrow[]{\rm uniform}$ & 70 &  37.2 & $-$ 2.9, + 2.4 & $-$ 8.30, + 7.81\\
Gas velocity dispersion ($\sigma$; \kms) &   0.0& $\xrightarrow[]{\rm uniform}$ &20.0 &    12.9 & $-$  $\pm$0.9 &$-$  2.7, +  3.6\\
\\
Nuisance parameters: &&&&&&\ \\\hline
Centre X offset (\arcsec) &  $-$1.0& $\xrightarrow[]{\rm uniform}$ &  1.0 &    0.0 & $\pm$0.01 & $\pm$0.02\\
Centre Y offset (\arcsec) &  $-$1.0& $\xrightarrow[]{\rm uniform}$ &  1.0 &     0.0 & $\pm$0.01 &$\pm$0.02\\
Centre velocity offset (\kms) & $-$20.0& $\xrightarrow[]{\rm uniform}$ & 20.0 &    $-$ 0.5 & $-$  1.2, +  1.5 & $-$  3.8, +  4.1\\ \hline
\end{tabular*}
\parbox[t]{0.8\textwidth}{ \textit{Notes:} Column 1 lists the fitted model parameters, while Column 2 lists the prior for each. $N$($\mu$,$\sigma$) indicates a Gaussian prior with a given mean ($\mu$) and standard deviation ($\sigma$). Where the prior is listed as uniform, this is uniform in linear space (or in logarithmic space for the IMBH mass only). The posterior distribution of each parameter is quantified in the third to fifth columns (see also Fig. 3). The X, Y and velocity offset nuisance parameters are defined relative to the International Celestial Reference System (ICRS) position 01$^{\rm h}$09$^{\rm m}$27\coordsec001, +35$^{\circ}$43$^{\prime}$04\farc942, and the systemic (barycentric) velocity of the galaxy V$_{\rm bary}$=$-$53 \kms.}
\end{center}
\label{fittable}
\end{table*}

\subsection{Uncertainties}
\label{uncerts}
In this Section we briefly touch upon the main uncertainties that could affect the IMBH mass derived in this work.

Firstly, given that we do not resolve the Keplerian increase of the circular velocity around the BH {(see Fig.~\ref{modelpvds})}, our mass measurement is crucially dependent on the stellar mass model and its mass normalisation factor ($\Gamma$), that together with the molecular gas itself set the contribution of luminous matter to the observed gas kinematics. This model was constructed from multi-band \textit{HST} images, and carefully fitted pixel-by-pixel by  \cite{2017ApJ...836..237N} to correct for stellar population gradients and dust obscuration. The mass normalisation factor used here is that which provided the best fit to the stellar kinematics in Section \ref{stellar_model} (and we incorporate the uncertainty of this measurement into our prior). Despite this, if the model were to be substantially biased, it would clearly have an effect on our derived BH mass. For instance, we can completely remove the need for an IMBH in NGC\,404 if $\Gamma>1.15$. Using the molecular gas data alone with a flat prior on $\Gamma$ shows that, while possible, such a value is not the most likely with $\Delta\chi^2\approx5$ (implying it deviates from the best fit by $\gtsimeq$2$\sigma$). This same $\Gamma$ would, however, be a 4.3$\sigma$ deviation from that found using stellar kinematics. The combination of these two datasets is thus crucial to allow us to accurately constrain the BH mass in NGC\,404.

Secondly, the velocity dispersion {(and the resulting vertical extension of the gas disc/torus)} provides several additional sources of uncertainty. The first of these arises because we have assumed that the velocity dispersion is constant as a function of radius. If instead the velocity dispersion were to change substantially, it could bias our best-fitting BH mass. To test this we re-ran our best-fitting model, allowing for a linear gradient in velocity dispersion. The best-fitting velocity dispersion gradient was consistent with zero, suggesting that a constant velocity dispersion within the nuclear molecular gas disc is reasonable. In addition, our models do not take into account the additional dynamical support of the molecular gas due to its velocity dispersion. Due to the relatively low total rotation velocity ($V$) of NGC\,404, this `asymmetric drift' correction is likely to be more important than in other systems with molecular gas BH mass measurements. We estimate the magnitude of this effect using the formalism described in Section 6.5 of \cite{2006MNRAS.366.1050C}. We assume that the random motions in the gas disc are nearly equal in the radial ($\sigma_{\rm r}$) and vertical ($\sigma_{\rm z}$) directions, while the ratio of the velocity dispersion in the radial and azimuthal directions ($\sigma_{\rm r}/\sigma_{\rm \phi}$) can be calculated under the epicyclic approximation, which is valid as long as $\sigma_{r}\ll V$. In our best-fitting model $|V|$ is always $>58$\,\kms\ in the central disc of NGC\,404 (the presence of the IMBH prevents the velocity decreasing further), and thus $V/\sigma\gtsimeq5$ everywhere. This method can thus at least provide an approximate treatment of the asymmetric drift. Given these assumptions, we can use Equation 12 of \cite{2006MNRAS.366.1050C} to calculate the asymmetric drift correction, that is then $<5\%$ everywhere within the central disc of NGC\,404. We thus conclude that asymmetric drift uncertainties are small compared to the other error sources, even in this low-mass object.

{A related uncertainty arises from the assumption that the molecular material in the disc/torus of this object is distributed in a geometrically thin disc. If this material were instead to be geometrically thick this could bias the BH mass we estimate. To investigate this possibility we ran a further fitting procedure where the thickness of the disc was allowed to vary. Specifically the density as a function of height above the disc falls off following a Gaussian distribution, with the scale-height (the standard deviation of the Gaussian) left as a free parameter. A full description of this process is given in Appendix \ref{diskthick}. Assuming the molecular gas disc/torus in the central parts of this object is geometrically thick does not change the best-fitting BH mass in a statistically significant way (i.e the best-fitting BH mass is consistent with that found in Section \ref{fitting} within the uncertainties). We are, however, able to set a 1$\sigma$ limit to the Gaussian scale-height in the vertical direction of $z_{\rm h}$<1.3 pc, and thus a limit to the geometric thickness of this molecular disc/torus at its outer edge ($r$=0\farc25 or 3.7\,pc) of $z_{\rm h}$/$r<0.35$. This result is consistent with simple theoretical expectations (see Appendix \ref{diskthick}).  As such we conclude that the geometric thickness of the molecular disc/torus at the centre of NGC\,404 is small, and not significantly biasing our derived BH mass.}

\begin{figure*}
\begin{center}
\includegraphics[width=0.95\textwidth,angle=0,clip,trim=0.0cm 0.0cm 0cm 0.0cm]{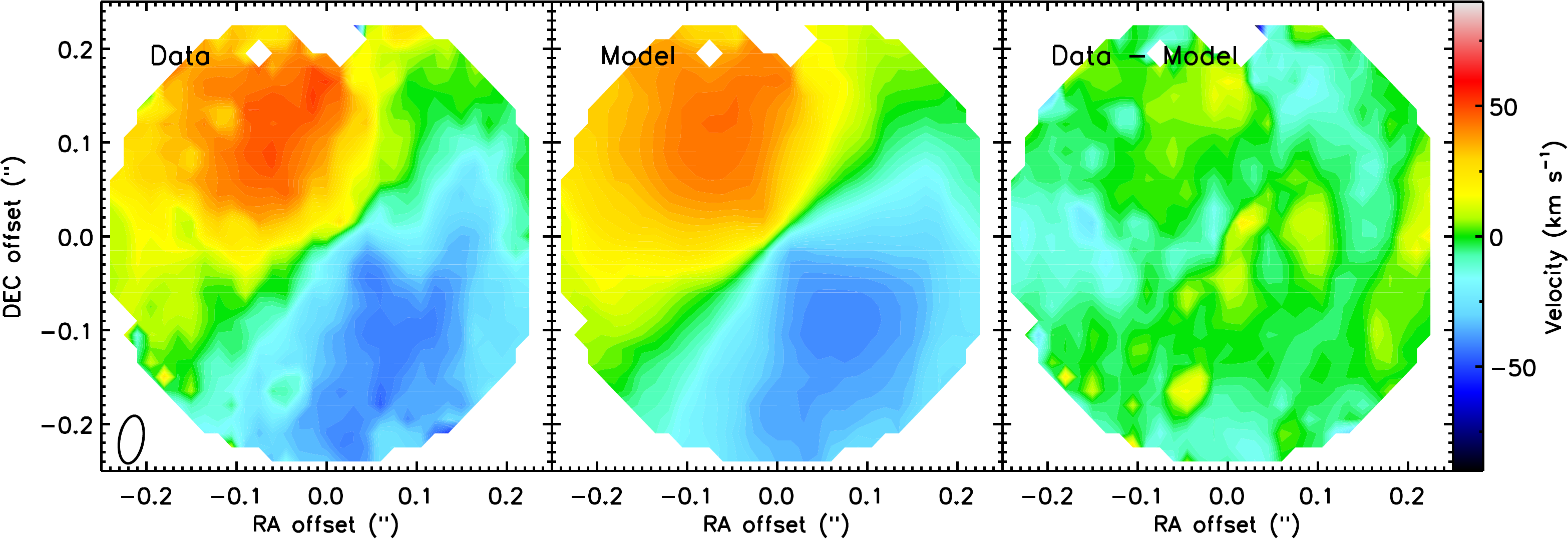}
\caption{\textit{Left and centre panels:} Moment one maps of the CO(2-1) emission of the central disc ($<$0\farc25 or $\approx$3.7\,pc) of NGC\,404, extracted from our observed data cube (left panel) and best-fitting model data cube (central panel). The synthesised beam (0\farc051\,$\times$\,0\farc026 or 0.54 pc$^2$) is shown in the bottom-left corner of the left panel. \textit{Right panel:} Residuals between the data and best-fitting model. The residuals are small (with a standard deviation of $\approx$7 \kms) and do not show any obvious spatial correlation, suggesting our fit is good and no significant non-circular motion is present in the cold molecular gas disc.  }
\label{vresid}
 \end{center}
 \end{figure*}

Thirdly, we assume that the molecular gas in the central disc of NGC\,404 is moving on circular orbits. This assumption is violated in the $\approx$2300\,K molecular gas probed by H$_2$ ro-vibrational lines \citep{2010ApJ...714..713S,2017ApJ...836..237N}, but we do not find evidence of significant non-circular motions in the cold gas from our ALMA observations. Our best-fitting model with purely circular motions provides a good fit to the data, as shown in Figure \ref{vresid}. The residuals around the best fit have a low amplitude and no obvious spatial structure. Any low level non-circular motions that may be present are thus unlikely to have significantly affected our BH measurement. 

Fourthly, as shown above, we know that the molecular gas mass contributes non-negligibly to the total mass of NGC\,404 at the radii we study, and thus our BH mass estimate depends on the assumed CO-to-H$_2$ conversion factor ($\alpha_{\rm CO}$). Here we have assumed the molecular gas has a standard, Galactic $\alpha_{\rm CO}$. This is motivated by the high metallicity of this object ($\approx$0.8\,Z$_{\odot}$ in its outer parts, and thus presumably more metal rich in its centre; \citealt{2013ApJ...772L..23B}). However, if the true $\alpha_{\rm CO}$ of NGC\,404 were to be significantly different, or vary radially, it could bias our BH mass estimate. 
Luckily, however, the two methods to estimate the BH mass we use here (stellar and gas kinematics) respond quite differently to the presence of the molecular gas mass.  Our molecular gas kinematic analysis assumes the material moves in circular orbits, and thus the motion of any given parcel of gas is primarily sensitive to (and primarily constrain) the mass enclosed within its orbit. Given that the majority of the molecular material is located at large radii, this makes our molecular gas measurement fairly insensitive to $\alpha_{\rm CO}$ (the molecular gas in the central disc is essentially a massless kinematic tracer). Indeed, allowing $\alpha_{\rm CO}$ to vary over a large range (0.1\,$< \alpha_{\rm CO} <$\,10\,\msun\,(K\,\kms\,pc$^2$)$^{-1}$) in our fitting procedure does not significantly affect the best-fitting BH mass. 
The stellar kinematic modelling, on the other hand, is conducted over a large field of view (and stellar orbits explore a larger range of radii), making it much more sensitive to the assumed $\alpha_{\rm CO}$.
Allowing a changing $\alpha_{\rm CO}$ in that analysis alters the derived BH mass by an order of magnitude. The fact that the two analyses agree well on the central BH mass when using a Galactic $\alpha_{\rm CO}$ thus suggests that this is a reasonable assumption for NGC\,404. 

Finally, all BH mass estimates are systematically affected by the distance ($D$) they assume to their target galaxy (with $M_{\rm BH} \propto D$). For NGC\,404 we use a tip of the red giant branch distance from \cite{2002A&A...389..812K}, that has an uncertainty of 12\%. Here then, as with all other BH mass measurements, the distance-related systematic uncertainty on the BH mass is significant.

\section{Discussion}
\label{discuss}

\subsection{Gas morphology}
\label{gasmorph}
The morphology of the molecular gas in NGC\,404 is complex. As discussed in detail above, a central disc/torus of CO(2-1) emission is present, which peaks at a radius of about 0\farc16 (2.3\,pc), and extends our to 0\farc25 (3.7\,pc). The disc itself is rotating around the IMBH, but is surrounded by kinematically-distinct lopsided emission that appears connected to an arm-like feature. This arm joins to an incomplete (pseudo)-ring of emission, with a radius of $\approx$2\farc7 (40\,pc), that corresponds well to dust features identified in absorption against the stellar continuum in Figure  \ref{ALMA_HST_over}. Scattered molecular clouds exist beyond this radius, whose filamentary distribution mirrors that seen in \hi\ at much larger radii \citep{2004AJ....128...89D}. 

The H$\alpha$+[\nii] emission detected by \textit{HST} anti-correlates with the presence of molecular gas (right panel of Fig. \ref{ALMA_HST_over}), appearing to fill in many of the missing portions of this molecular (pseudo)-ring. This may be due to extinction in the dense molecular gas preventing detection of optical line emission where they are coincident, to star-formation feedback dissociating/removing gas in some parts of the ring, or to temporal stochasticity in the star-formation process at these small spatial scales. 

Interestingly the Milky Way's central molecular zone (CMZ) also hosts a lopsided `ring' of molecular emission with star formation concentrated on one side \citep[e.g.][]{2011ApJ...735L..33M}. NGC\,404 could thus host an interesting analogue to the CMZ that we are able to study face on. We do note, however, that some of the H$\alpha$ near the centre of NGC\,404 may be due to an ionisation cone around the active galactic nucleus (AGN; \citealt{2017ApJ...845...50N}) or a recent supernova remnant \citep{2018ApJ...866...79B}. Disentangling the true mechanism ionising the gas would require high spatial resolution observations of other nebula lines and is beyond the scope of this work.

The kinematics of the arm and ring features outside of the nuclear disc are complex. As shown in the first moment map in Figure \ref{ALMAkinmoments}, the kinematic position angle appears to significantly change at radii $>$0\farc7 ($>$10\,pc), and the red/blue-shifts of the individual molecular clouds detected around the ring do not vary smoothly with position. This may be because we are seeing a truly chaotic distribution of clouds that do not lie in a settled disc. Alternatively, this outer material could be significantly more face-on than the disc around the galaxy nucleus (i.e. an inclination warp is present, as suggested by our kinematic modelling in Section \ref{model}), and thus the line-of-sight velocities we measure could be dominated by out-of-plane motions. The dynamical time at a radius of $\approx$2\farc7 (40\,pc), where the incomplete (pseudo)-ring of emission lies, is $\approx5.8$ Myr so in a flattened axisymmetric potential the warped disc we observe would be expected to relax quickly. The fact that we observe it suggests again that the potential of NGC\,404 is approximately spherical at these radii.

The arm-like feature that extends between the outer (pseudo)-ring and the central disc of molecular emission appears to link the two, and it may be funnelling gas from the outer regions inwards (potentially fuelling the AGN). Our maps provide evidence that the arm hosts strong streaming motions, as the second moment (shown in Fig. \ref{ALMAkinmoments} and \ref{ALMAmoments_zoom}) reveals large line-widths along it. The fact that this arm-like feature is not symmetric could be because gas inflow is mainly stochastic, or it may point to a recent disturbance in the system.

The properties of the individual molecular clouds observed here will be discussed further in a future work (Liu et al., in prep.).  

  \begin{figure*}
\begin{center}
\includegraphics[height=6cm,angle=0,clip,trim=0.0cm -0.1cm 0cm 0.0cm]{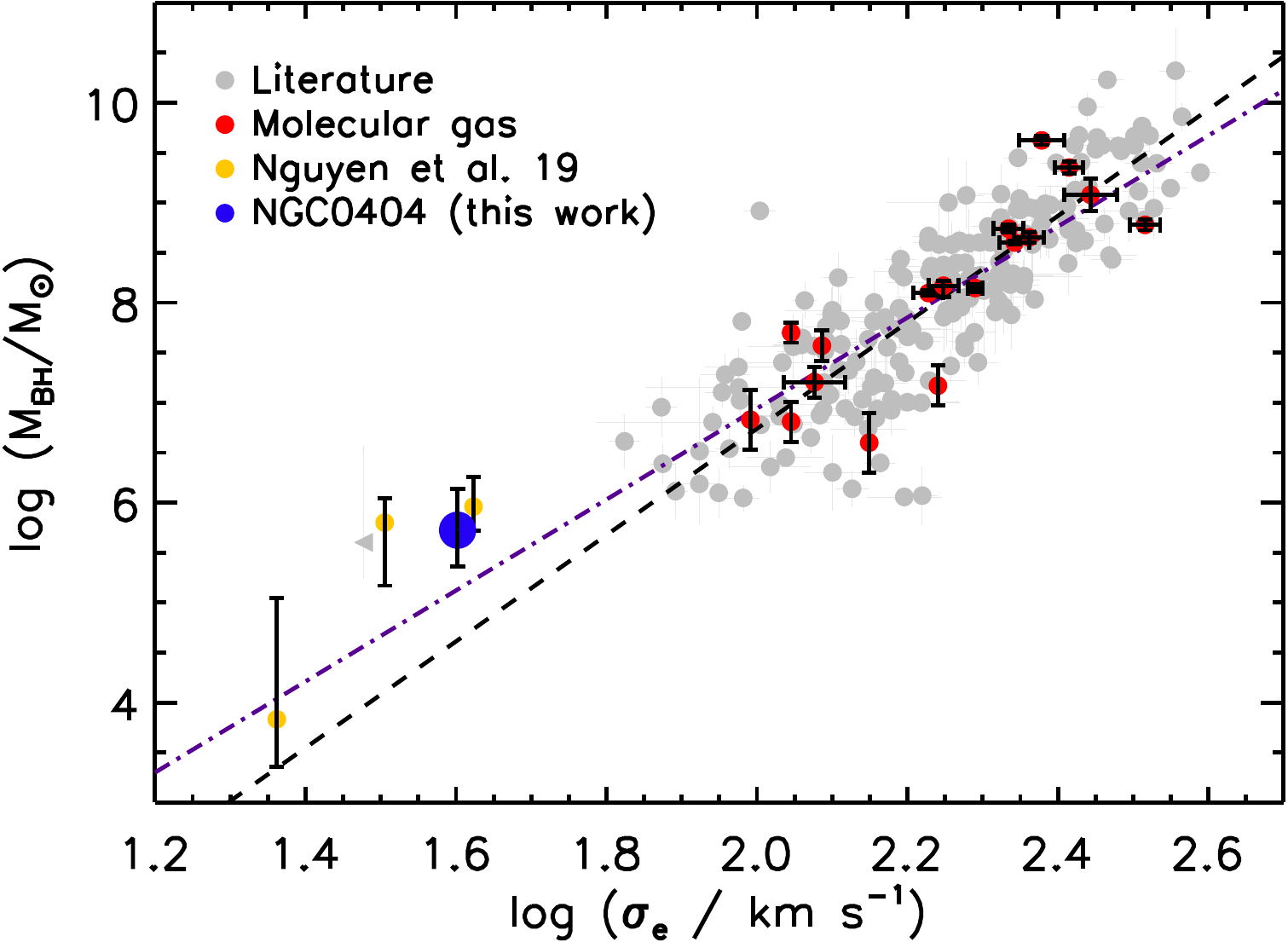}
\includegraphics[height=6cm,angle=0,clip,trim=1.7cm 0.0cm 0cm 0.0cm]{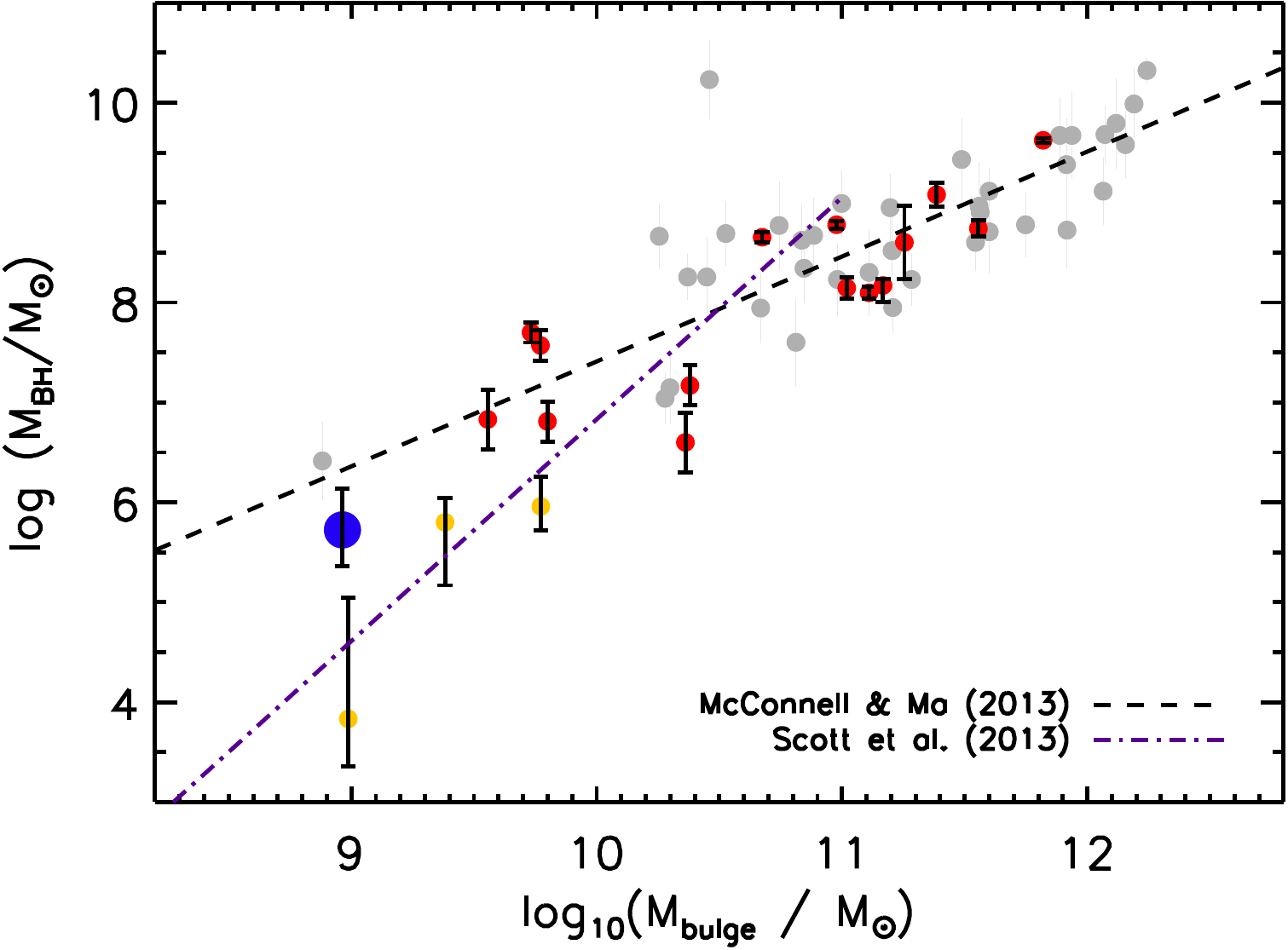}
\caption{\textit{Left panel:} $M_{\rm BH}$ -- $\sigma_{\rm e}$ relation (grey data points and black dashed line) from the compilation of \protect \cite{2016ApJ...831..134V}. We show the BH mass measured for NGC\,404 here as a large blue data point, and highlight other measurements using the molecular gas technique with red data points (\protect \citealt{2013Natur.494..328D}; \protect \citealt{2015ApJ...806...39O}; \protect \citealt{2016ApJ...822L..28B,2016ApJ...823...51B}; \protect \citealt{2017MNRAS.468.4663O}; \protect \citealt{2017MNRAS.468.4675D,2018MNRAS.473.3818D};  \protect \citealt{2019A&A...623A..79C}; \protect \citealt{2019MNRAS.485.4359S}; \protect \citealt{2019ApJ...883..193N}; \protect \citealt{2019MNRAS.490..319N}; \protect \citealt{2019ApJ...881...10B}; \citealt{2019arXiv190203813N}). In yellow we show the BH masses of low-mass galaxies measured in \protect \cite{2019ApJ...872..104N}. We also show the best fitting relation (for all galaxies, including upper limits in the IMBH regime) from \protect \cite{2019arXiv191109678G} as a purple dot-dashed line. \textit{Right panel:} $M_{\rm BH}$ -- $M_{\rm bulge}$ relation from the compilation of \protect \cite{2013ApJ...764..184M} (black dashed line) and \protect \cite{2013ApJ...768...76S} (purple dot-dashed line). Symbols are as in the left figure. Bulge masses for objects with molecular gas-estimated BH masses are taken from \protect \cite{2013ApJ...764..184M}, \protect \cite{2013MNRAS.432.1768K}, \protect \cite{2015ApJS..219....4S} and \protect \cite{2016MNRAS.457..320S}. In both panels, our measurement of the BH mass of NGC\,404 is reasonably consistent with the best-fitting relation(s).}
\label{N404_msig}
 \end{center}
 \end{figure*}
\subsection{Continuum emission}
\label{cont_discuss}

As discussed above, and shown in Figure \ref{ALMAcont_moments}, we detect spatially resolved 1\,mm (237\,GHz) continuum emission from the centre of NGC\,404, and some additional point sources. 

The central resolved source (labelled A in Fig. \ref{ALMAcont_moments}) is coincident with the molecular gas disc and arm at the core of NGC\,404. Given its morphology, tracing the one-armed spiral connecting the central molecular disc to the (pseudo-)ring at larger radii, we consider it unlikely that this emission is from a background source. This emission was not detected in the previous ALMA observations presented in \cite{2017ApJ...845...50N}; it may have been resolved out.  This  feature has two distinct flux density maxima, one around the nucleus of the galaxy (coincident with the small radio jet identified over a range of radio frequencies; e.g. \citealt{2017ApJ...845...50N}) and one at the base of the one-armed spiral feature. We note that this central source is detected in a higher resolution naturally-weighted continuum map made without $uv$-tapering from our combined dataset (at a resolution of 0\farc15$\times$0\farc13). At this resolution only the brightest areas around the nucleus, and the feature around the base of the arm are detected.

Various physical mechanisms could be causing this extended continuum emission. The centrally-concentrated flux around the black hole overlaps with hot dust emission ($\approx$950\,K) observed in the near infrared \citep{2010ApJ...714..713S,2019arXiv190907323D}, the extended H$\alpha$ source visible in Figure \ref{ALMA_HST_over}, and the small radio jet. We consider it likely that the majority of this emission is from the Rayleigh-Jeans tail of the dust continuum, as its flux is far higher than the pure-synchrotron extrapolation of the jet spectral energy distribution \citep{2017ApJ...845...50N}, but we cannot rule out the presence of a very flat-spectrum radio core. 

The second continuum maximum at the base of the molecular `arm' could again contain contributions from several possible physical mechanisms.  This is a location where one might expect shock heating to be important, as the gas flows inward along this arm towards the centre. Indeed, Figure \ref{ALMAkinmoments} shows an elevated molecular gas velocity dispersion along this feature. $C$- or $J$-shocks in the molecular medium can lead to continuum emission, and if the material flowing into the shock moves fast enough, or is very dense, the predicted continuum flux density can match that found for this source \citep[e.g.][]{1987ApJ...322..275M}. Such shocks should, however, be even brighter at 15 GHz, but this feature was not detected in the $Ku$-band data of \cite{2017ApJ...845...50N}. We thus conclude that this extended continuum feature is again likely dominated by dust continuum emission, with possible small contributions from synchrotron in the jet region and hydrodynamic shocks along the arm.

Three other continuum sources are significantly detected in our continuum image. One (labelled B in Fig. \ref{ALMAcont_moments})  lies to the east, overlapping with a $\approx$1\farc3 (20\,pc) long molecular filament, while another (labelled C in Fig. \ref{ALMAcont_moments}) lies to the south-west, at the base of a filamentary spur of molecular emission. The final source (labelled D in Fig. \ref{ALMAcont_moments}) is located well to the north, away from the regions with detected molecular gas. Once again, it is not clear what powers the emission in these sources. 
They are associated neither with radio emission at 15 GHz, nor significant H$\alpha$ emission. While H$\alpha$ emission can easily be hidden due to dust, the lack of radio emission suggests that the emission is not free-free from H\textsc{\small II} regions nor synchrotron from supernova remnants.

Another possibility is that these three sources are the most extreme star-forming clumps/cores in NGC\,404, that stand out from the diffuse dust background that may be resolved out. If their continuum luminosities are due to Rayleigh-Jeans emission from dust, then assuming a temperature of 50\,K and a canonical gas-to-dust ratio of 100, we find that the total masses of the clumps would be at least 6$\times$10$^4$\,\msun, making them similar to the most massive clumps/cores in the Milky Way \citep{2018MNRAS.473.1059U}. Additional continuum data at higher frequencies would be required to confirm this hypothesis. 

The remaining possibility is that some/all of these continuum sources are background galaxies. Based on the model of \cite{2013ApJ...768...21C}, and given our sensitivity, we would expect to detect $\approx$3-4 high-$z$ sources in our primary beam at 3$\sigma$ significance. As such, this is a viable explanation of these emission features, but the fact that 3/4 of these sources are clustered in the inner $4$\arcsec$\times4$\arcsec\ is suspicious. Deeper and/or higher frequency observations will be needed to definitively understand the diverse continuum emission from NGC\,404.

 \subsection{Intermediate-mass black hole}
 \label{imbh}
 Our modelling reveals the presence of a dark object in NGC\,404, of a mass of $\approx5\times10^5$\,\msun, this from both stellar kinematic and molecular gas kinematic modelling. 
 The radius of the sphere-of-influence of this BH (${R}_{\rm SOI}$) can be calculated using the standard equation
 
 \begin{equation}
{R}_{\rm SOI}= \frac{GM_{\rm BH}}{\sigma_{\rm e}^2},
\end{equation}
where $G$ is the gravitational constant and $\sigma_{\rm e}$ the stellar velocity dispersion. In NGC\,404 this equates to 0\farc1 (or $\approx$1.4\,pc). This ${R}_{\rm SOI}$ is of similar size to the point-spread function of our optical data ($\approx$0\farc12), and is well resolved by our high-resolution molecular gas data with a synthesised beam of 0\farc051\,$\times$\,0\farc026 (or $\approx$0.54 pc$^2$). 
 
Crucially, to bring our stellar and molecular gas measurements into agreement we needed to include the contribution of the molecular gas to the gravitational potential of this object. While the molecular material does not dominate the potential at any radius, it does provide a significant contribution towards the outer edge of the central molecular disc (see Fig. \ref{N404_massplot}). Including its effect decreases the derived mass mismatch factor ($\Gamma$), and increases the black hole mass required compared to the analysis in \cite{2017ApJ...836..237N}, bringing these two measurements into agreement. It also brings the stellar kinematic measurement into better agreement with the hot-H$_2$ kinematic measurement of \cite{2017ApJ...836..237N}. This highlights the need to accurately determine the mass and distribution of molecular material around intermediate-mass BHs when attempting to measure their mass, even when using stellar kinematics. 
 
 In Figure \ref{N404_msig} we show our measurement of the BH mass in NGC\,404 as a blue data point on the $M_{\rm BH}$--$\sigma_{\rm e}$ relations of \cite{2016ApJ...831..134V,2019arXiv191109678G} and on the $M_{\rm BH}$--bulge mass ($M_{\rm bulge}$) relation of \cite{2013ApJ...764..184M}. We highlight other measurements using the molecular gas technique with red data points. Bulge masses for objects with molecular gas estimated BH masses are taken from \protect \cite{2013ApJ...764..184M}, \cite{2013MNRAS.432.1768K}, \cite{2015ApJS..219....4S} and \protect \cite{2016MNRAS.457..320S}, while velocity dispersions are as quoted in the mass measurement papers. The mass of the IMBH in NGC\,404 agrees reasonably well with the extrapolations of the relations presented in these works, being slightly above the $M_{\rm BH}$-$\sigma_e$ and slightly below the $M_{\rm BH}$-M$_{\rm bulge}$ relation. Obtaining more measurements of low-mass BHs is clearly crucial to better anchor these relations.

\subsection{Stellar mass normalisation}

Our kinematic modelling uses a model of the stellar mass of NGC\,404 created from \textit{HST} photometry using pixel-by-pixel colour to mass-to-light ratio relations, carefully normalised using stellar population estimates of the mass-to-light ratios based on Space Telescope Imaging Spectrograph (STIS) spectra assuming a \cite{2003PASP..115..763C} IMF.
As such, this model should provide an excellent measurement of the stellar contribution to the potential.  

Here we also included in the fit a stellar mass normalisation factor, that should be approximately unity assuming no systematic effect is present. However, as shown above, our stellar kinematic modelling requires a significantly lower stellar mass normalisation ($\Gamma\approx0.6$). This suggests the stellar population is significantly less massive than expected assuming a \cite{2003PASP..115..763C} IMF. 

\begin{figure}
\begin{center}
\includegraphics[width=0.45\textwidth,angle=0,clip,trim=0.0cm 0.0cm 0cm 0.0cm]{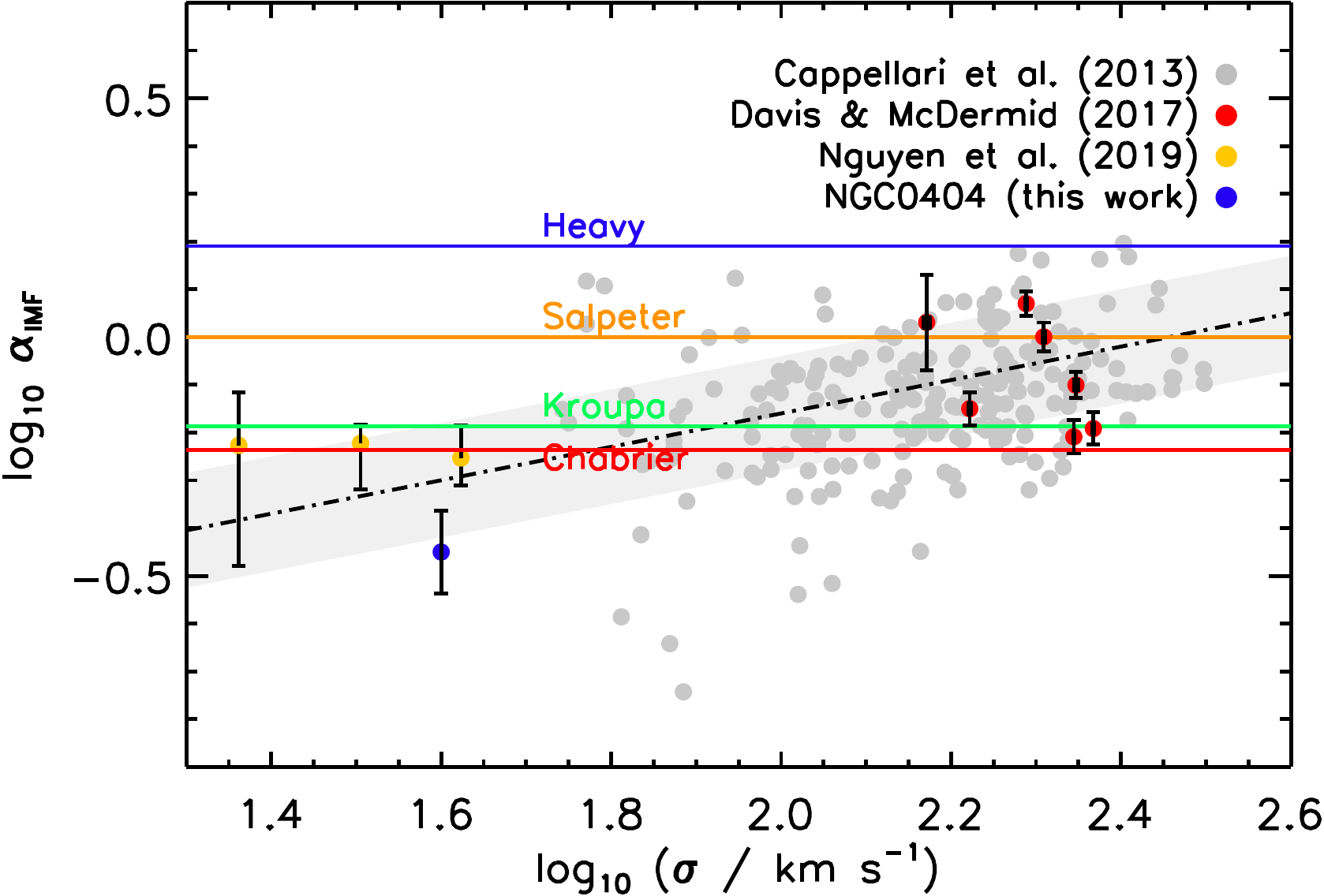}
\caption{The IMF mismatch parameter ($\alpha_{\rm IMF}$) plotted against the stellar velocity dispersion ($\sigma$) measured within the same aperture. Our measurement in NGC\,404 is shown in blue, while measurements from \protect \cite{2012Natur.484..485C},  \protect  \cite{2017MNRAS.464..453D} and  \protect  \cite{2019ApJ...872..104N} are shown in grey, red and orange, respectively. Also shown are solid coloured lines that denote the $\alpha_{\rm IMF}$ parameter of a heavy, Salpeter, Kroupa and Chabrier IMF. A black dot-dashed line denotes the best fit correlation of \protect \cite{2012Natur.484..485C,2013MNRAS.432.1709C}, and the grey shaded region around it denotes the reported scatter around that relation of 0.12 dex. Our estimate of the IMF mismatch in NGC\,404 is marginally consistent with the extrapolation of the best-fitting relation.}
\label{N404_imf}
 \end{center}
 \end{figure}

The cause of this change is difficult to establish. One possibility is that the IMF of NGC\,404 is significantly lighter than assumed. Significant variations have been observed in the IMFs of massive early-type galaxies (see e.g. \citealt{2010Natur.468..940V,2012Natur.484..485C,2015ApJ...806L..31M}), and the extrapolations of at least some of the trends reported in the literature would support the idea of a lower IMF normalisation in this dwarf elliptical galaxy.
For instance, some authors \citep[e.g.][]{2012Natur.484..485C} suggest that the velocity dispersion of an ETG is the best predictor of its IMF.  Figure \ref{N404_imf} shows the IMF measurements from that work (as tabulated in \citealt{2013MNRAS.432.1862C}) and those of \cite{2017MNRAS.464..453D} and \cite{2019ApJ...872..104N}, plotted against stellar velocity dispersions measured within the same apertures. Shown in blue is our estimate of the IMF mismatch parameter $\alpha_{\rm IMF}$ in NGC\,404, which is our stellar mass normalisation factor $\Gamma$ re-normalised to a \cite{1955ApJ...121..161S} IMF via $\alpha_{\rm IMF}=0.58\,\Gamma$ (where this proportionality constant comes from ratio of the integral of these IMFs). While the $\alpha_{\rm IMF}$ values derived here (and those of \citealt{2019ApJ...872..104N}) are only valid for the nuclear star clusters of their parent galaxy (rather than for the entire galaxy), our measurement for NGC\,404 is consistent with the extrapolation of the best-fitting relation of \cite{2012Natur.484..485C,2013MNRAS.432.1709C}. Interestingly, NGC\,404 has a high metallicity and thus, if real, the low stellar-mass normalisation of NGC\,404 would be less consistent with metallicity -- IMF relations \citep[e.g. from][]{2015ApJ...806L..31M}. This suggests that further work on the IMF properties of dwarf elliptical galaxies could help shed light on the drivers of IMF variations in the early-type galaxy population \citep[see e.g.][]{2016MNRAS.463.2819M}.

Having said that, a changing IMF is not the only possible explanation of our low derived $\Gamma$. Another possibility is that the star formation history (SFH) of NGC404 has not been correctly estimated by \citet{2017ApJ...836..237N}, who used a non-parametric approach fitting combinations of models with ages of 1, 10, 50, 100, 300, 600, 1000, 2500, 5000, 10,000 and 13,000 Myr.  The low $\Gamma$ obtained here could be explained if the best SFH fit has too many old stars (or not enough young stars).  \citet{2017ApJ...836..237N} found an unusual stellar population in the nucleus, with a dominant 1~Gyr old population.  This is an age where asymptotic giant branch stars contribute significantly to the luminosity, and it is possible these are not accurately modelled (which could result in a mis-estimation of the SFH).  We note that in the nuclear star clusters of three other galaxies studied by \cite{2019ApJ...872..104N} (and shown in orange in Figure \ref{N404_imf}) without significant molecular gas (and thus star formation), the $\Gamma$ values are all consistent with a Chabrier IMF, albeit with relatively large uncertainties. Yet another possibility is that we have overestimated the mass of gas present in the nucleus of NGC\,404 (e.g. by assuming an incorrect $\alpha_{\rm CO}$). Despite the differing sensitivities of our stellar and gas BH mass measurement to this quantity (see Section \ref{uncerts}), this possibility cannot be ruled out using our current data alone. We therefore suggest that the low $\Gamma$ value should be treated with caution.

\section{Conclusions}
\label{conclude}

In this work we presented observations of the molecular ISM in the dwarf elliptical galaxy NGC\,404 at an unprecedented linear resolution of $\approx$0.5\,pc. We used these observations, in conjunction with Gemini observations of the stellar kinematics, to estimate the mass of the intermediate-mass black hole at the heart of this object. 

Our ALMA observations revealed a central disc/torus of molecular gas which peaks in surface density at a radius of $\approx$0\farc16 (2.3\,pc) {and has an exponential scale-height smaller than $1.3$\,pc}. This disc is inclined to the line of sight and is clearly rotating around the IMBH. It is surrounded by a kinematically-distinct lopsided arm/spiral feature, connecting to a nearly face-on incomplete (pseudo)-ring of emission with a radius of 2\farc7 ($\approx$40\,pc). Scattered molecular clouds also exist beyond this radius.

Continuum emission is detected from the central parts of NGC\,404 (tracing the central disc and lopsided arm) and from several other locations at larger radii. The extended central continuum source is likely dominated by dust continuum emission, with possible small contributions from synchrotron emission from the small jet in the centre, and hydrodynamic shocks along the arm. The outer sources could be the most extreme molecular star-forming clumps in this object or an over-density of background galaxies. 

By combining the molecular gas data with constraints on galaxy parameters from stellar kinematic modelling we were able to constrain the mass of the IMBH in NGC\,404 using two independent techniques. We showed here that, once we include the contribution of the molecular gas to the gravitational potential, the observed stellar kinematics require a black hole mass of 5.1$^{+14}_{-4.5}\times$10$^5$ \msun\ (at the 99\% CL). We were also able to estimate the mass of the IMBH in NGC\,404 directly from the molecular gas kinematics, utilising a forward modelling approach in a Bayesian framework to fit the observed data cube. 
 We again clearly detect the presence of a dark object, with a mass of 5.5$^{+4.1}_{-3.8}\times$10$^5$ \msun\ (again at the 99\% CL), in good agreement with our stellar kinematic measurement. In both cases the BH mass inferred is larger than that previously estimated using the stellar kinematics alone. This is because the molecular material in this object contributes non-negligibly to the potential, systematically biasing the determination of the BH mass. This highlights the need to accurately determine the mass and distribution of all dynamically important components around low mass BH's when attempting to measure their mass.

Our derived black hole mass for NGC404 is broadly consistent with extrapolations from the $M_{\rm BH}$-$\sigma_e$ and $M_{\rm BH}$-$M_{\rm bulge}$ relations. Drawing strong conclusions about the distribution of IMBH masses and how they correlate with galaxy properties will, however, require larger samples. ALMA is poised to contribute significantly to this, both through providing maps of the dynamically important molecular material, and by directly tracing its rotation around the black holes themselves.

 \vspace{0.5cm}
\noindent \textbf{Acknowledgments}

TAD acknowledges support from a Science and Technology Facilities Council Ernest Rutherford Fellowship and through grant ST/S00033X/1. TAD thanks Michael Anderson, Mattia Negrello, Andrew Rigby and Elizabeth Watkins for helpful discussions, and the anonymous referee whose suggestions improved this paper. Research by ACS is supported by NSF grant AST-1350389. Basic research in radio astronomy at the U.S. Naval Research Laboratory is supported by 6.1 Base Funding. MB was supported by the consolidated grants `Astrophysics at Oxford' ST/N000919/1 and ST/K00106X/1 from the UK Research Council. Research by AJB is supported by NSF grant AST-1614212. TGW acknowledges funding from the European Research Council (ERC) under the European Union's Horizon 2020 research and innovation programme (grant agreement No. 694343). 

\noindent  This paper makes use of the following ALMA data:\begin{itemize}
\item  ADS/JAO.ALMA\#2015.1.00597.S
\item ADS/JAO.ALMA\#2017.1.00572.S
\item ADS/JAO.ALMA\#2017.1.00907.S
\end{itemize}

\noindent ALMA is a partnership of ESO (representing its member states), NSF (USA) and NINS (Japan), together with NRC (Canada), MOST and ASIAA (Taiwan), and KASI (Republic of Korea), in cooperation with the Republic of Chile. The Joint ALMA Observatory is operated by ESO, AUI/NRAO and NAOJ. The National Radio Astronomy Observatory is a facility of the National Science Foundation operated under cooperative agreement by Associated Universities, Inc.

This paper is also based, in part, on observations made with the NASA/ESA Hubble Space Telescope, and obtained from the Hubble Legacy Archive, which is a collaboration between the Space Telescope Science Institute (STScI/NASA), the Space Telescope European Coordinating Facility (ST-ECF/ESA) and the Canadian Astronomy Data Centre (CADC/NRC/CSA).

\bsp
\bibliographystyle{mnras}
\bibliography{bibN404.bib}
\bibdata{bibN404.bib}
\bibstyle{mnras}

\label{lastpage}

\appendix
\section{Molecular gas kinematic fitting including disc thickness}
\label{diskthick}

{As discussed in Section \ref{uncerts} we are able to adapt our fitting procedure to study the effect of the assumed thickness of the molecular gas disc/torus in the centre of NGC\,404. We add a free parameter for the disc thickness (denoted $z_{\rm h}$) to the model described in Section \ref{model}, which we assume is radially constant. The molecular `cloudlets' that make up our \textsc{skySampler} model are given randomly allocated offsets in the $z$-direction drawn from a Gaussian distribution centred around zero, and with a standard deviation of $z_{\rm h}$. This three-dimensional distribution of gas cloudlets is then input to the \textsc{KinMS} simulation routines, which calculate the potential of the resulting molecular gas distribution, and create mock observations which can be directly compared to the data as part of our MCMC fitting procedure.}

\begin{figure*}
\begin{center}
\includegraphics[width=0.85\textwidth,angle=0,clip,trim=0.0cm 0.0cm 0cm 0.0cm]{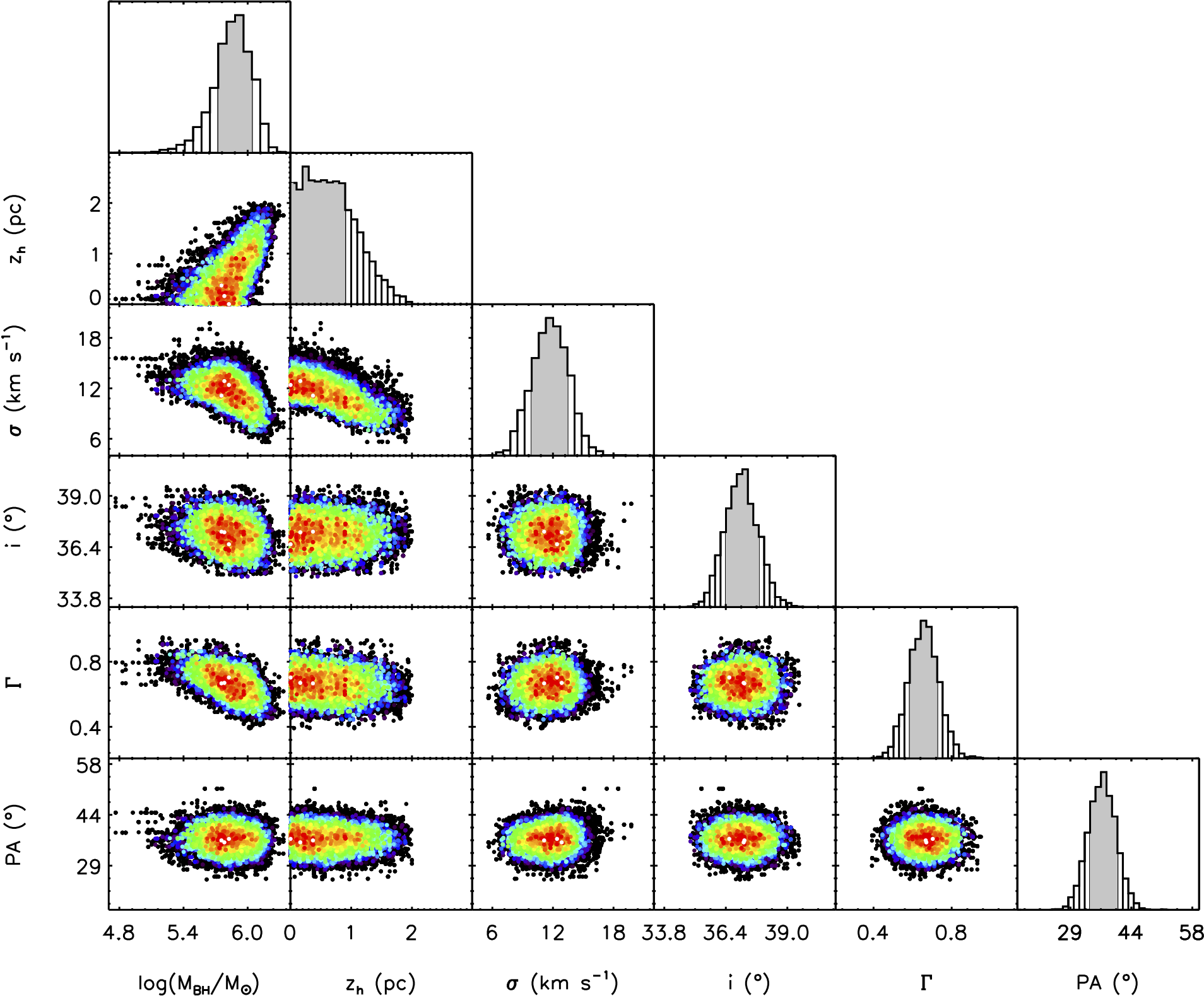}
\caption{{Results of kinematic fitting to the molecular gas in NGC\,404, as Figure \ref{triangleplot}, but including the effect of disc thickness ($z_{\rm h}$). Assuming the molecular gas disc/torus in the central parts of this object (with a radius $r=3.7$\,pc) is geometrically thick does not change our best-fitting BH mass significantly. We are, however, able to set a 1$\sigma$ limit to the vertical scale-height of $z_{\rm h}$<1.3 pc.}}
\label{N404_diskthick}
 \end{center}
 \end{figure*}

{The results of this analysis are shown in Figure \ref{N404_diskthick}. We obtain an upper limit on the scale-height $z_{\rm h}$ itself, constraining $z_h<1.3$\,pc at the 1$\sigma$ confidence level. While the BH mass we derive does have a dependence on $z_{\rm h}$ (with thicker discs requiring a larger best-fitting BH mass), this change is not significant within the uncertainties. We thus do not expect the thickness of the molecular torus in this object to be significantly affecting the main results of this paper. }

{ The limit we derive for the molecular scale-height in NGC\,404 is fully consistent with expectations from simple theoretical models. 
Following \cite{1992pavi.book.....S}, the vertical gas density profile in a non self-gravitating gas disc rotating in a spherically symmetric potential will have a Gaussian form at any radius $r$ (consistent with the assumption made for the vertical distribution in our modelling), where the scale height $z_{\rm h}$ is given by}
\begin{equation}
z_h=\sqrt{2}\left(\frac{\sigma_{\mathrm{z}}}{V_{\phi}}\right)r,
\end{equation}
{where $\sigma_{\mathrm{z}}$ is the gas velocity dispersion in the vertical direction and $V_{\phi}$ is the component of the velocity in the azimuthal direction. As described above $V_{\phi}/\sigma_z\gtsimeq5$ everywhere in the central regions of NGC\,404, and thus at the outer edge of the molecular disc/torus we would predict $z_h\ltsimeq0.8$\,pc, consistent with our derived upper limit.}

{Alternatively one can consider the case of a disc dominated by its own self gravity, where the vertical scale height is given by}
\begin{equation}
z_h=\frac{\sigma_{\mathrm{z}}^2}{\pi G \Sigma_{\rm gas}},
\end{equation}
{where $\Sigma_{\rm gas}$ is the gas surface density \citep{1942ApJ....95..329S}. Inputting the measured values of these quantities for NGC\,404 ($\Sigma_{\rm gas}\approx$\,10,000\msun\,pc$^{-2}$, $\sigma_z= 12.9$\,\kms) gives an estimate of $z_h=1.2$\,pc, again consistent with our derived upper limit. We note that the vertical gas density distribution in this case should have a sech$^2$(z) form, and thus this scale-height is not directly comparable with that derived above. Their similarity does, however, suggest our upper limit to the scale-height of the disc/torus in NGC\,404 is reasonable. }

\end{document}